\documentclass[%
reprint,
superscriptaddress,
 amsmath,amssymb,
 aps, prl, twocolumn,
]{revtex4}

\usepackage{graphicx}
\usepackage{dcolumn}
\usepackage{bm}
\usepackage{enumitem}
\usepackage{placeins}
\usepackage[usenames,dvipsnames]{color}
\usepackage{hyperref}

\begin{document}

\title{Finding discrete symmetry groups via Machine Learning}

\author{Pablo Calvo-Barl\'es}
\affiliation{Instituto de Nanociencia y Materiales de Arag\'on (INMA), CSIC-Universidad de Zaragoza, 50009 Zaragoza, Spain\looseness=-1}
\affiliation{Departamento de F\'isica de la Materia Condensada, Universidad de Zaragoza, 50009 Zaragoza, Spain\looseness=-1}

\author{Sergio G. Rodrigo}
\affiliation{Instituto de Nanociencia y Materiales de Arag\'on (INMA), CSIC-Universidad de Zaragoza, 50009 Zaragoza, Spain\looseness=-1}
\affiliation{Departamento de F\'isica Aplicada, Universidad de Zaragoza, 50009 Zaragoza, Spain\looseness=-1}

\author{Eduardo S\'anchez-Burillo}
\affiliation{PredictLand S.L., 50001 Zaragoza, Spain\looseness=-1}

\author{Luis Mart\'in-Moreno}
\email[]{lmm@unizar.es}
\affiliation{Instituto de Nanociencia y Materiales de Arag\'on (INMA), CSIC-Universidad de Zaragoza, 50009 Zaragoza, Spain\looseness=-1}
\affiliation{Departamento de F\'isica de la Materia Condensada, Universidad de Zaragoza, 50009 Zaragoza, Spain\looseness=-1}

\date{\today}

\begin{abstract}
We introduce a machine-learning approach (denoted Symmetry Seeker Neural Network) capable of automatically discovering discrete symmetry groups in physical systems. This method identifies the finite set of parameter transformations that preserve the system's physical properties. Remarkably, the method accomplishes this without prior knowledge of the system's symmetry or the mathematical relationships between parameters and properties. Demonstrating its versatility, we showcase examples from mathematics, nanophotonics, and quantum chemistry.
\end{abstract}

\maketitle



Symmetries play an essential role in the understanding of physical theories. For example, continuous symmetries lead to conservation laws \cite{Noether1918}, and discrete symmetries (such as parity or time reversal) are key in studying spectral degeneracies in quantum mechanical systems \cite{Cohen-Tannoudji:101367}.

Unveiling hidden symmetries is thus of immense interest in many fields of physics.  Until recently, this task relied solely on human intuition. However,  recent advances in Machine Learning (ML), and especially in Neural Networks (NN) \cite{LeCun_2015,Cybenko1989,Hornik1991,nielsen2015neural,Carleo_2019}, have opened up new and efficient ways of analyzing data. NNs have already been successfully employed in problems such as learning invariants and conservation laws of dynamical systems \cite{Liu_2022,Liu_2021,https://doi.org/10.48550/arxiv.2305.19525,Ha_2021,Wetzel_2020,https://doi.org/10.48550/arxiv.1906.01563,https://doi.org/10.48550/arxiv.2003.04630,https://doi.org/10.48550/arxiv.2208.14995}; finding coordinate transformations that reveal hidden symmetries \cite{Liu_2022_Hid_Sym,https://doi.org/10.48550/arxiv.1906.04645}; learning Lie algebras \cite{Craven_2022,Forestano_2023,Krippendorf_2020,https://doi.org/10.48550/arxiv.2302.00236,NEURIPS2021_148148d6}; discovering transformations that preserve the statistical distribution of data sets \cite{Desai_2022,https://doi.org/10.48550/arxiv.2302.00236}; recognizing symmetries from NN embedding layers \cite{Krippendorf_2020,Barenboim_2021} and classifying whether pairs of events are related by symmetry or not \cite{Wetzel_2020,Decelle_2019}.

\begin{figure}
    \centering
    \includegraphics[scale=0.053]{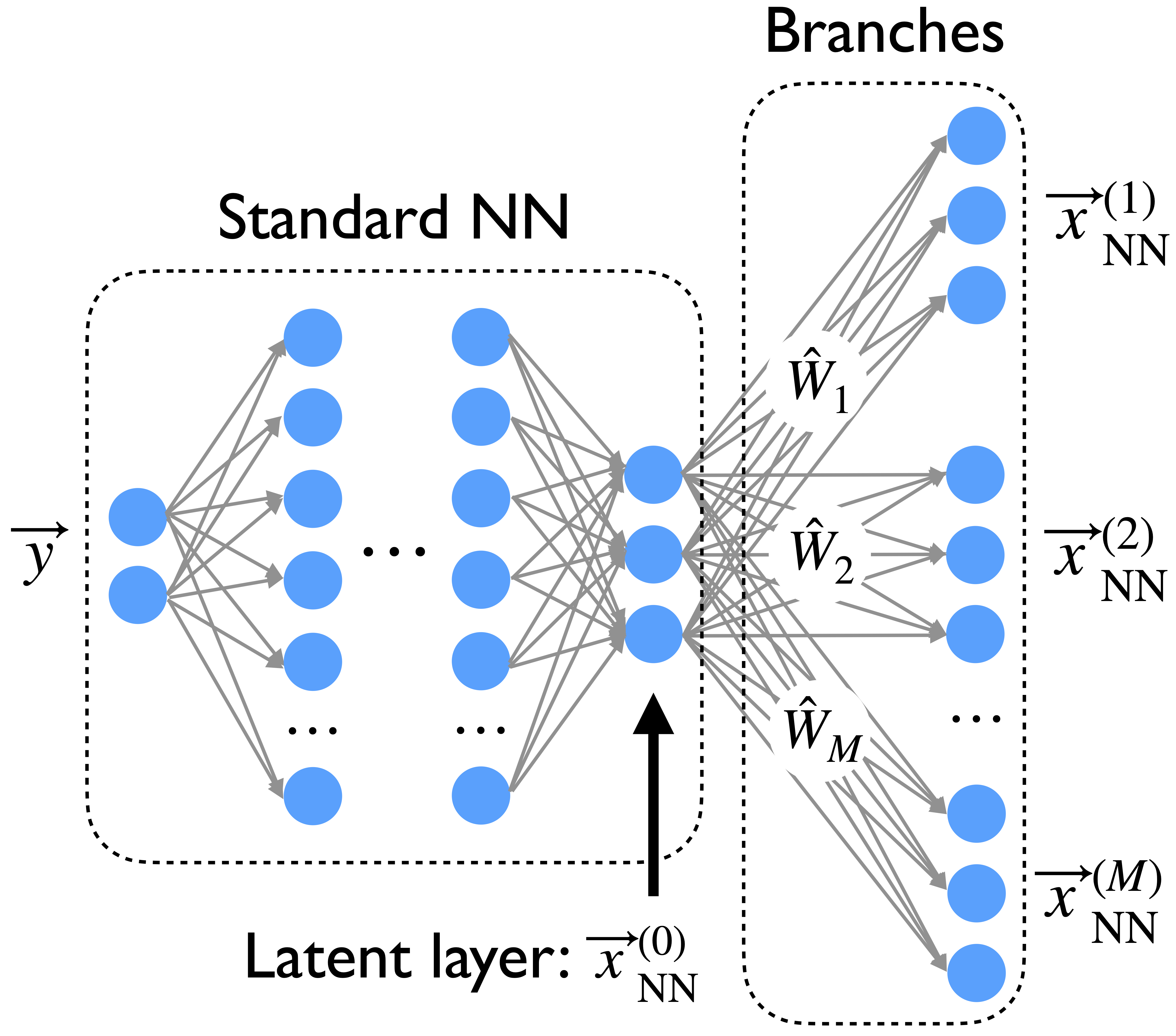}
    \caption{Schematic representation of the SSNN: Magnitudes $\vec{y}$ are fed into a standard NN, producing a vector $\vec{x}_{\mathrm{NN}}^{(0)}$.  This NN is complemented with a set of $M$ branches.  The $\alpha$-th branch performs a linear transformation $\hat{W}_{\alpha} \vec{x}_{\mathrm{NN}}^{(0)}$, providing a prediction $\vec{x}_{\mathrm{NN}}^{(\alpha)}$ for the physical parameters.  Without the branches, the presence of symmetries would lead to inaccurate results. In contrast, the SSNN, trained with the customized loss function given by \eqref{eq:loss}, accurately predicts the parameters using available data only.  After training, the branches $\hat{W}_{\alpha}$ provide a representation of the system's symmetry group.}
    \label{fig:FIG_1}
\end{figure}

A typical situation not covered by the previous studies is when the parameters that control the physical magnitudes of a system have a discrete set of symmetry transformations that leave these magnitudes invariant. 
To find these transformations, we introduce a NN architecture (which we denote as Symmetry Seeker Neural Network, SSNN) capable of discovering the matrix representation of all symmetry group elements using only a dataset of physical parameters and corresponding measurable magnitudes. Importantly, our method requires no prior knowledge of the symmetry group or the mathematical relations between parameters and physical properties. Additionally, the SSNN can be employed in inverse design problems with symmetry-related multivalued solutions.

The text is structured as follows: we first introduce the SSNN algorithm and demonstrate its effectiveness on a simple mathematical problem. Next, we apply the method to examples from nanophotonics and quantum chemistry.


\begin{figure}
    \centering
    \includegraphics[scale=0.057]{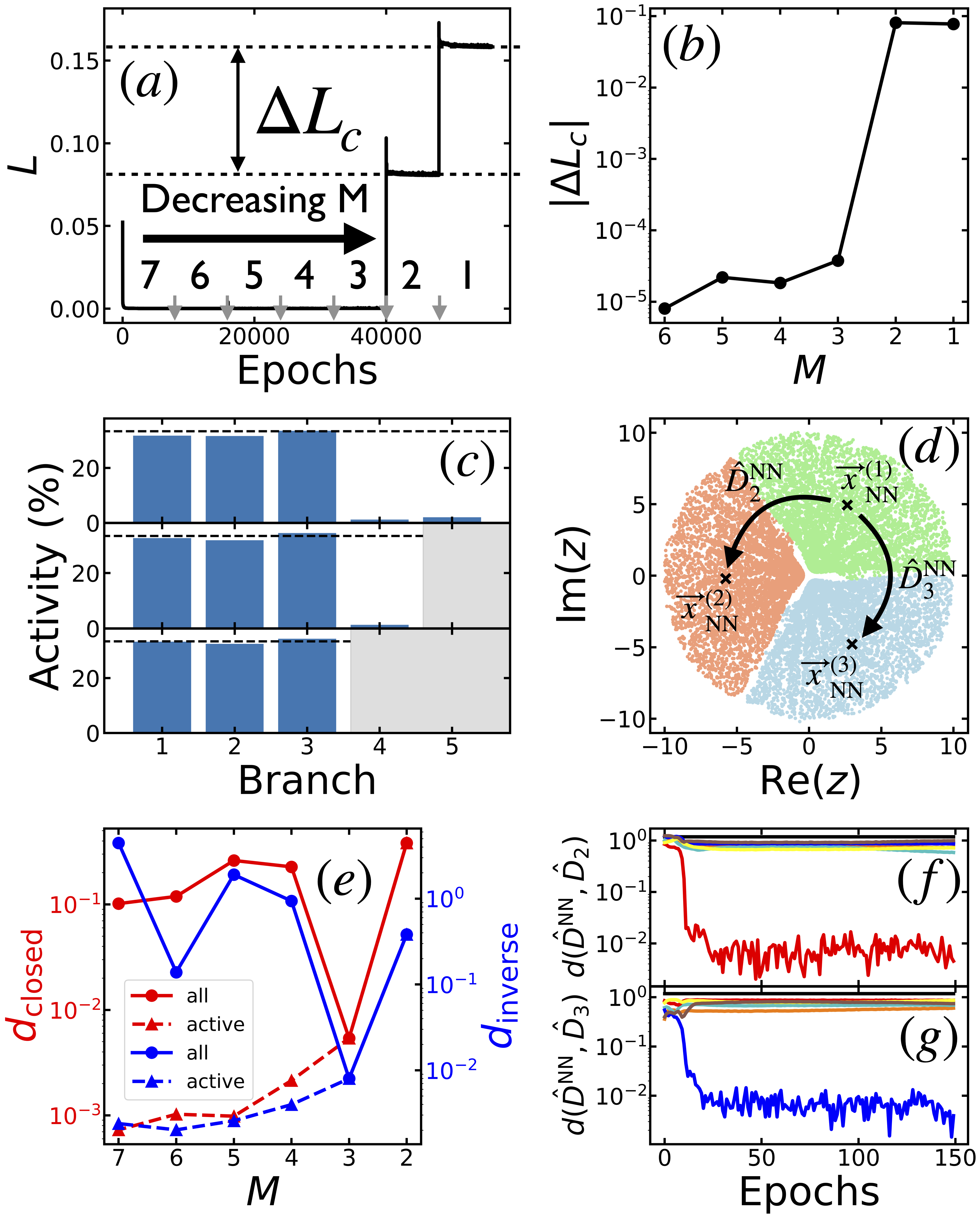}
    \caption{Iterative SSNN algorithm for $w = z^3$. (a) Learning curve with grey arrows indicating the epoch of branch removal. (b) Loss jump dependence with the number of branches. (c) $A_{\alpha}$ for $\alpha = 1,...,M$ at $M=5$ (above), $M=4$ (middle), and $M=3$ (below), with the black dashed line representing $1/K = 1/3$. Grey panels at $M=4,3$ indicate removed branches. (d) Predictions of the training data set from SSNN with $M=3$. Three points highlighted by black crosses represent predictions $\vec{x}_{\mathrm{NN}}^{(\alpha)}(\vec{y})$ from a single input data related by $\hat{D}_{\alpha}^{\mathrm{NN}}$. (e) Group metrics showing $d_{\mathrm{closed}}$ (red curves) and $d_{\mathrm{inverse}}$ (blue curves) for all branches (solid lines) and active branches (dashed lines). (f) Evolution of the distance between the $\hat{D}_{\alpha}^{\mathrm{NN}}$ and the ground truth matrix  $\hat{D}_{2}$ during the first training stage ($M = 7$). (g) Same as (f) for $\hat{D}_{3}$}
    \label{fig:FIG_2}
\end{figure}

To define the problem in general terms, we consider a function $\vec{y}(\vec{x})$, where the ``magnitudes'' $\vec{y} \in \mathbb{R}^m$ depend on the ``parameters'' $\vec{x} \in \mathbb{R}^n$. 
The ``system'' may present discrete symmetries,  so symmetry-related parameters lead to the same magnitudes.  If so, the symmetry transformations form a finite group $G = \{ d_1,...,d_{K}\}$ of order $K$ with a $n \times n$ matrix representation $\{ \hat{D}_1,...,\hat{D}_K \}$ that satisfies:
\begin{equation}\label{eq:symmetry}
\vec{y}(\vec{x}) = \vec{y}(\hat{D}_{\alpha}\vec{x}), \; \forall \vec{x}, \alpha.
\end{equation}
We assume that all information about the system is represented by $N$ training samples $\{ ( \vec{x}_i , \vec{y}_i ) \}_{i=1}^N$ (where $\vec{y}_i = \vec{y}(\vec{x}_i)$).  The goal is to find an accurate prediction $\{ \hat{D}^{\mathrm{NN}}_{\alpha}\}_{\alpha = 1}^K$ of the unknown ground truth symmetry matrices $\{ \hat{D}_{\alpha}\}_{\alpha = 1}^K$.  

For that, we define the SSNN architecture schematically represented in Fig. \ref{fig:FIG_1}. It is a NN that receives as input $\vec{y}$ and outputs $M$ predictions of $\{ \vec{x}^{(\alpha)}_{\mathrm{NN}} \}_{\alpha=1}^{M}$ (we will use $i$ to index the training samples and $\alpha, \beta$ to index the symmetry matrices). The SSNN is designed to find the inverse function $\vec{y}^{-1}$, which can be multivalued due to symmetry. It consists of two parts: the first part is a neural network that produces an $n$-dimensional vector $\vec{x}^{(0)}_{\mathrm{NN}}$. The last layer of this part is called the ``latent" layer. The second part consists of $M$ trainable linear transformations $\hat{W}_{\alpha}$, called branches, which generate $M$ vectors $\vec{x}^{(\alpha)}_{\mathrm{NN}} = \hat{W}_{\alpha} \vec{x}^{(0)}_{\mathrm{NN}}$.

The SSNN's performance is guided by the defined loss function, given by:
\begin{equation}\label{eq:loss}
    L = \frac{1}{2N} \sum_i \min_{\alpha} L_i^{(\alpha)},
\end{equation}
where $L_i^{(\alpha)} = || \vec{x}_i - \vec{x}^{(\alpha)}_{\mathrm{NN}}(\vec{y}_i)||^2$
%
is the mean square error of the prediction $\vec{x}^{(\alpha)}_{\mathrm{NN}}(\vec{y}_i)$ with respect to the training sample $\vec{x}_i$.
For each training data point $i$, only the branch with the lowest $L_i^{(\alpha)}$ (referred to as the ``winner" branch for point $i$) contributes to the loss.  To gauge the contribution of each branch in the training process, we introduce the concept of branch activity, denoted as $A_{\alpha} = N_{\alpha}/N$, where $N_{\alpha}$ is the number of times the branch $\alpha$ had winner predictions across the entire training data set.

It is worth noting that a related structure to the SSNN has been proposed for inverse design problems \cite{Zhang_2018}. However, the SSNN incorporates two crucial modifications: first, the latent layer has the dimension of the parameter space; second, the branches are defined by linear transformations. These changes allow the branches to be related to the symmetry representation matrices, as demonstrated in the subsequent analysis.

To find the symmetry group, we use an iterative method for training the SSNN, which involves removing the least active branch in each step.  The SSNN is initiated with a large enough number of branches $M_{\mathrm{max}}$ (which must be larger than the group order $K$). The SSNN is trained until the loss $L$ converges and the branch with the lowest $A_{\alpha}$ is pruned.  This training plus pruning process is repeated until the network has only one branch, and the converged loss $L_c (M)$ is recorded at each stage. We refer to each iteration as a training stage.

To illustrate our method, we apply it to a mathematical example: the complex cube root $w=z^3$, where $z$ and $w$ are complex numbers.  This problem exhibits a symmetry group of rotations of order $K=3$ in the complex plane, as $z^3 = (e^{i2\pi / 3} z)^3 = (e^{i4\pi / 3} z)^3 , \; \forall z$.  In the representation of complex numbers $\vec{x} = (\mathrm{Re}(z),\mathrm{Im}(z))^T$ and $\vec{y} = (\mathrm{Re}(w),\mathrm{Im}(w))^T$, where $T$ states for transposition,   the $2 \times 2$ symmetry matrices are:
\begin{equation}\label{eq:dalpha}
    \hat{D}_{\alpha} = \begin{pmatrix}
                        \cos{\theta_{\alpha}} & \sin{\theta_{\alpha}} \\
                        -\sin{\theta_{\alpha}} & \cos{\theta_{\alpha}}
                        \end{pmatrix},
\end{equation}
with $\theta_{\alpha} = 2\pi (\alpha-1)/3 $, with $\alpha = 1,2,3$. 

We train the model by randomly selecting 6000 values of $z$ with $|z|<10$. The main results are presented here, while the details of the training/validation data set generation, SSNN hyperparameters, and other examples for different root powers can be found in the Supplemental Material (SM) \cite{suppmat}.

The training process is illustrated in Fig. \ref{fig:FIG_2}. In Fig. \ref{fig:FIG_2}(a), we show the evolution of $L$ as $M$ decreases during the training epochs, starting from an initial value of $M_{\mathrm{max}} = 7$.  Figure \ref{fig:FIG_2}(b) shows the jump in converged loss at consecutive stages, defined as $\Delta L_c (M) = L_c (M+1) - L_c (M)$. These plots strongly indicate a group order of $K = 3$.  As $\vec{y}^{-1}$ is a $K$-fold symmetry-multivalued function, $M \geq K$ predictions are required to fit all possible $\vec{y}^{-1}$ solutions and achieve $L_c \simeq 0$.  However, when $M < K$, there are not enough branches to predict all possible solutions, leading to improper parameter predictions for certain data points and resulting in $L_c$ being orders of magnitude higher.  This is apparent in Fig. \ref{fig:FIG_2}(b), where $\Delta L_c(M)$ jumps from $10^{-5}$ to $10^{-1}$ after $M=3$, providing the first indicator that the group order is $K=3$.

The activities of the branches further confirm this. Fig. \ref{fig:FIG_2}(c) represents $\{ A_{\alpha} \}_{\alpha = 1}^M$ at the end of the stages $M = 5,4,3$.  In all cases, at the end of each training stage, there are precisely $K$ active branches equally having $A_{\alpha} \simeq 1/3 = 1/K$ and $M-K$ inactive branches having $A_{\alpha} \simeq 0$.  This happens because during the training process, the activity of $K$ branches is sufficient to provide all solutions to $\vec{y}^{-1}$. In contrast, the remaining branches become less active, contributing less often to the loss and ultimately becoming detached from the training refinement.

Let's examine the actions of the active branches. Fig. \ref{fig:FIG_2}(d) shows the SSNN predictions $\vec{x}^{(\alpha)}_{\mathrm{NN}}(\vec{y}_i)$ ($\alpha = 1,...,M$) for every training sample $\vec{y}_i$ at the end of the stage $M=3$.  Each branch provides predictions confined to a distinct irreducible region of the symmetry group.  These irreducible regions, all of equal size, do not overlap with each other,  and together they encompass the entire parameter space. Additional plots of predictions from inactive branches for $M>K$, along with a visual representation of the activities $A_{\alpha}$, are provided in the SM \cite{suppmat}.

We observe that each irreducible region transforms into another region when 3-fold rotations are applied, indicating a connection between the symmetry transformations of the problem and the predictions among different branches.  Thus,  we can extract the predicted symmetry matrices $\hat{D}_{\alpha}^{\mathrm{NN}}$ from the active branches $\hat{W}_{\alpha}$. 
However, a degeneracy issue arises: $\hat{D}_{\alpha} \vec{x}^{(0)}_{\mathrm{NN}} = (\hat{D}_{\alpha}\hat{A}^{-1})(\hat{A}\vec{x}^{(0)}_{\mathrm{NN}})$ for any invertible matrix $\hat{A}$, resulting in infinite sets of  $ \{ \hat{W}_{\alpha} \}$ that yield the same loss.  To resolve this, we exploit the fact that any symmetry group includes the identity as an element. Hence, among the branches identified by the SSNN, we choose a reference matrix $\hat{W}_{\mathrm{ref}}$ (we take $\hat{W}_{\mathrm{ref}} = \hat{W}_1$) and compute the predicted symmetry matrices as:
\begin{equation} 
    \hat{D}^{\mathrm{NN}}_{\alpha} = \hat{W}_{\alpha} \hat{W}_{\mathrm{ref}}^{-1}.
\end{equation}

Apart from including the identity, a group must satisfy two other fundamental properties \cite{ARFKEN2013815}:  closure under multiplication ($\forall \hat{D}_{\alpha}, \hat{D}_{\beta} \in G, \; \exists \hat{D}_{\gamma} \in G$ such that $\hat{D}_{\alpha}\hat{D}_{\beta} = \hat{D}_{\gamma}$), and containing the inverse of every group element ($\forall \hat{D}_{\alpha} \in G, \; \exists \hat{D}_{\gamma} \in G$ such that $\hat{D}_{\alpha}\hat{D}_{\gamma} = \hat{D}_{\gamma}\hat{D}_{\alpha} = \hat{1}$).  Since these two properties are not enforced in our calculation, we can employ them to validate the symmetry matrices discovery.  For this purpose, we define two group metrics as functions of $\{ \hat{D}_{\alpha}^{\mathrm{NN}} \}$ that quantify how well these matrices satisfy the group properties. For that, we first define a distance measure between two arbitrary matrices $\hat{A}$ and $\hat{B}$:
\begin{equation}
    d( \hat{A} , \hat{B} ) = \frac{1}{n^2} \sum_{k l} |a_{kl} - b_{kl}|,
\end{equation}
where $k$ and $l$ index the matrix elements and $n$ is the dimension of the matrices. The metric for determining whether the set $\{ \hat{D}^{\mathrm{NN}}_{\alpha} \}$ is closed under multiplication (``closed metric'') is defined as,
\begin{equation}
    d_{\mathrm{closed}} (\{ \hat{D}^{\mathrm{NN}}_{\alpha} \}) = \frac{1}{M^2} \sum_{\alpha \beta} \min_{\gamma} d( \hat{D}^{\mathrm{NN}}_{\alpha}\hat{D}^{\mathrm{NN}}_{\beta} , \hat{D}^{\mathrm{NN}}_{\gamma}),
\end{equation}
Similarly, the ``inverse metric" quantifies whether the set ${\hat{D}^{\mathrm{NN}}_{\alpha}}$ has an inverse belonging to the set:
\begin{equation}
    d_{\mathrm{inverse}}(\{ \hat{D}^{\mathrm{NN}}_{\alpha} \}) = \frac{1}{M} \sum_{\alpha} \min_{\gamma} d( \left[ \hat{D}^{\mathrm{NN}}_{\alpha}\right]^{-1}, \hat{D}^{\mathrm{NN}}_{\gamma} ).
\end{equation}
Clearly, the matrices $\{ \hat{D}^{\mathrm{NN}}_{\alpha} \}$ form a group if and only if $d_{\mathrm{closed}} = d_{\mathrm{inverse}} = 0$.

In Fig. \ref{fig:FIG_2}(e), we assess the discovery of the group using two different approaches for the group metrics. Firstly, we calculate these metrics considering all the predicted matrices, both from active and inactive branches. For $M = K = 3$, both the closed and inverse metrics approach zero, indicating that the matrices indeed form a group. However, when $M < K$, some predicted matrices lie outside the group after multiplication or inversion, leading to an increase in the group metrics. Similarly, for $M > K$, inactive branches are not sufficiently optimized, resulting in matrices that do not belong to the symmetry group, thus explaining the increase in the group metrics.

In the second approach, we only consider predicted matrices from the active branches when $M > K$, excluding the inactive branches. The closed and inverse metrics remain small for $M > K$, demonstrating that the active branches accurately predict the symmetry group matrices, even when $M$ exceeds the group order.

In this simple example, where we know the ground truth symmetry matrices $\{ \hat{D}_{\alpha} \}$, we can track the optimization of the predicted  $\hat{D}_{\alpha}^{\mathrm{NN}}$ throughout the epochs at the first training stage ($M=7$). Figure \ref{fig:FIG_2}(f) shows the distance between each predicted $\hat{D}_{\alpha}^{\mathrm{NN}}$ and $\hat{D}_2$ (see Eq. \eqref{eq:dalpha}). Figure \ref{fig:FIG_2}(g) does the same for $\hat{D}_3$ (we omit the curve for the identity transformation as $\hat{D}_{1}=\hat{D}^{\mathrm{NN}}_{1}$ by design). In each panel, one curve (corresponding to an active branch) gradually decays to a value that is orders of magnitude lower than that of the remaining curves. This confirms that the matrices ${ \hat{D}_{\alpha}^{\mathrm{NN}} }$ from active branches provide an excellent approximation to the ground truth $\{ \hat{D}_{\alpha} \}$.


\begin{figure}
    \centering
    \includegraphics[scale=0.043]{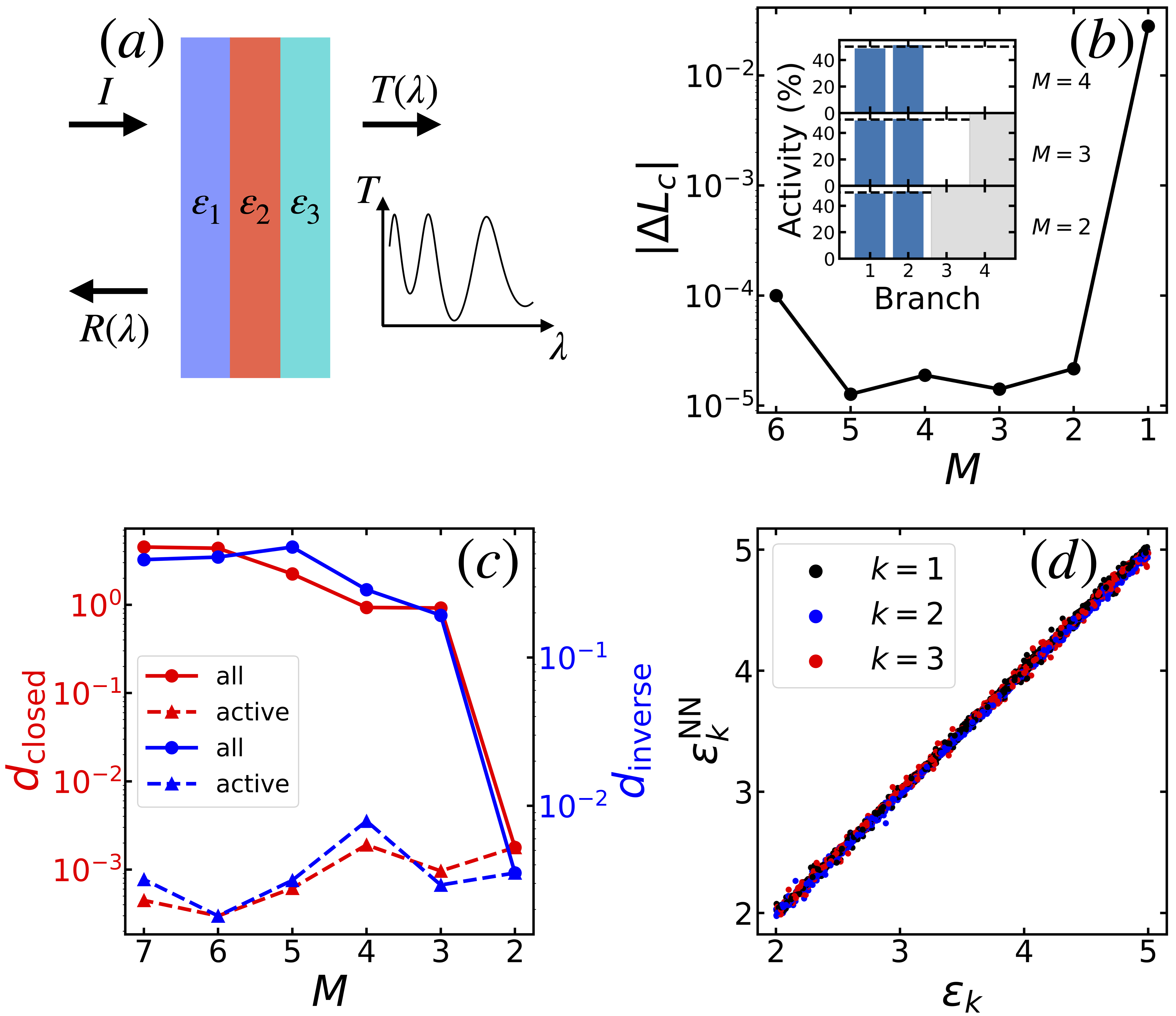}
    \caption{ Optical transmitted spectra symmetries in a multilayer stack:
(a) Schematic representation of the considered photonic system.
(b) Loss jump variation with the number of branches. The inset displays $A_{\alpha}$ for $\alpha = 1,...,M$ at three different stages: $M=4,3,2$. The black dashed line represents $1/K = 1/2$. The grey regions at $M=3,2$ indicate the removed branches.
(c) Group metrics depicted as a function of $M$.
(d) Dielectric constant predictions by the SSNN as a function of the true values of $\varepsilon_1$, $\varepsilon_2$, and $\varepsilon_3$. }
    \label{fig:FIG_3}
\end{figure}

Let us now consider the application of the SSNN to two physical examples. In the first one, we consider the dielectric multilayer stack represented in Fig. \ref{fig:FIG_3}(a). The system is composed of three layers, all with the same fixed width $d=250\; \mathrm{nm}$. Each layer has a dielectric constant $\varepsilon_i$ ($i = 1,2,3$), which we allow to have a value between 2 and 5.  
An electromagnetic plane wave impinges at normal incidence with a wavelength $\lambda$. By solving Maxwell's equations, we compute the transmission spectrum $T(\lambda)$ for a given choice of the material dielectric constants (see the SM \cite{suppmat} for further details). In this case, the mapping is between a set of dielectric constants $\vec{x} = (\varepsilon_1, \varepsilon_2,\varepsilon_3)^T$ with a transmission spectrum evaluated on a discretized set of wavelengths $\vec{y} = \left[ T(\lambda_1),...,T(\lambda_k),..., T(\lambda_m)\right]^T$.  For the calculations we use $m=100$ $\lambda$'s, equally distributed between $500 \; \mathrm{nm}$ and $900 \; \mathrm{nm}$.  The system's time reversal symmetry implies that the transmission spectrum remains invariant if the order of materials in the stack is inverted \cite{Newton1982}. In other words, the transmission spectrum for the set $(\varepsilon_1, \varepsilon_2, \varepsilon_3)^T$ is the same as for $(\varepsilon_3, \varepsilon_2, \varepsilon_1)^T$.  Consequently, the problem possesses a symmetry group of order $K=2$, consisting of the identity ($\hat{D}_1$) and the inversion transformation ($\hat{D}_2$).

Figure \ref{fig:FIG_3}(b) shows $\Delta L_c (M)$, when we initialize the SSNN with $M_{\mathrm{max}}= 7$ branches.  The jump of $\Delta L_c (M)$ in several orders of magnitude when $M<2$, indicates that the order of the group is $K=2$.  A further indication is found in the inset of that panel,  which shows two active branches with $A_{\alpha} \simeq 1/2$.  The closed and inverse metrics represented in Fig. \ref{fig:FIG_3}(c) demonstrate that the matrices corresponding to the active branches indeed form a group (in the SM \cite{suppmat} we show that the predicted matrix $\hat{D}_{2}^{\mathrm{NN}}$ converges to $\hat{D}_{2}$). 

Finally, we showcase the potential of the proposed method in inverse design. To achieve this, we employ the SSNN prediction from the winner branch denoted as $(\varepsilon_1^{\mathrm{NN}}, \varepsilon_2^{\mathrm{NN}}, \varepsilon_3^{\mathrm{NN}})$ for each data point.

Figure \ref{fig:FIG_3}(d) displays the comparison between the ground truth $\varepsilon_{k}$ and the corresponding $\varepsilon_{k}^{\mathrm{NN}}$ for all multilayer stacks in the validation set.  The results show an excellent agreement, confirming the method's effectiveness in providing inverse design solutions for all symmetry-related cases.

\begin{figure}
    \centering
    \includegraphics[scale=0.056]{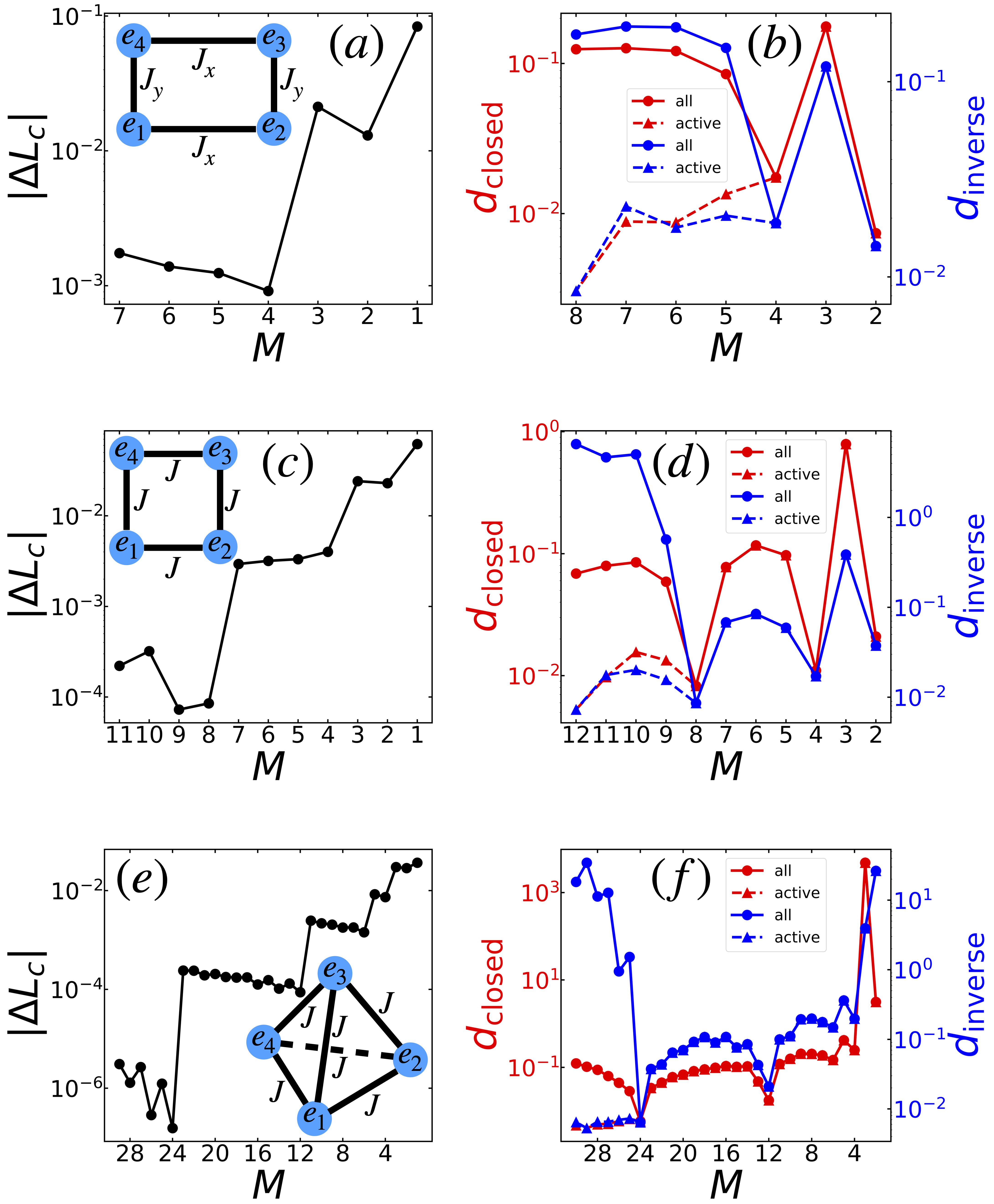}
    \caption{ Application of the SSNN to finding symmetries in the spectra of molecular systems. For each molecule (rectangular, square, and tetrahedral), two aspects are shown: the loss jump and the group metrics (for both all branches and the active ones) as a function of the number of branches. Figures (a-b), (c-d), and (e-f) correspond to the rectangular, square, and tetrahedral molecules, respectively. }
    \label{fig:FIG_4}
\end{figure}

Lastly, we apply our method to molecular systems in which discrete symmetry groups play an important role. We consider a tight-binding model of a molecule with 4 atoms, each having one orbital with atomic energy $e_i$, for $i=1,2,3,4$.  We assume that electron hopping only occurs between nearest-neighbor atoms, with a hopping constant $J$, which we take as the unit of energy.  The molecule's geometry defines the hopping terms (see the SM \cite{suppmat} for further details). The eigenvalues of the system's Hamiltonian are denoted as $E_i$, for $i=1,2,3,4$.  In this scenario, the mapping $\vec{y}$ associates a tuple of atomic energies, $\vec{x} = (e_1,e_2,e_3,e_4)^T$, with a tuple of molecular eigenvalues, $\vec{y} = (E_1,E_2,E_3,E_4)^T$. 
We explore three different molecular geometries: rectangular, square, and tetrahedral [see insets in Figs. \ref{fig:FIG_4}(a,c,e)]. 
For each molecular geometry, we generate 100000 ``molecules'' by ranging randomly choosing the values for $\{e_{i}\}$ and solving the Schr\"odinger equation to obtain the corresponding $\{E_{i}\}$. 
The symmetry transformations of these systems preserve the set of eigenvalues and include rotations, reflections, and improper rotations. The orders of these groups are $K = 4$ for the rectangle, $K = 8$ for the square, and $K = 24$ for the tetrahedron. In terms of the defined mapping $\vec{y}$, these transformations are manifest as permutations of the atomic energies $e_i$. For instance, in the case of the square molecule, a $\pi / 2$ rotation is given by the transformation $\hat{D}$ such that $\hat{D}(e_1,e_2,e_3,e_4)^T = (e_4,e_1,e_2,e_3)^T$. 

For each molecular geometry, we train a separate SSNN, starting with $M_{\mathrm{max}} = 8$ for the rectangle, $M_{\mathrm{max}} = 12$ for the square, and $M_{\mathrm{max}} = 30$ for the tetrahedron. The corresponding $\Delta L_c (M)$ is presented in Figures \ref{fig:FIG_4}(a,c,e), while Figures \ref{fig:FIG_4}(b,d,f) display the closed and inverse group metrics, considering both all branches and only the active ones. 
The presence of jumps in $\Delta L_c (M)$ and the behavior of the group metrics as $M$ changes clearly demonstrate that, in all cases, the SSNN successfully identifies the correct group order and symmetry matrices. This conclusion is further supported by the convergence of the predicted matrices to the ground truth matrices, as shown in the SM \cite{suppmat}. 

It is worth noting that, for each molecular geometry, Figure \ref{fig:FIG_4} reveals deep minima in both $d_{\mathrm{closed}}(M)$ and $d_{\mathrm{inverse}}(M)$ at values $M<K$. These minima correspond to a set of $\{ \hat{D}_{\alpha}^{\mathrm{NN}} \}$ that form a subgroup within the full system. For instance, when considering the tetrahedron, at $M = 12$ the set $\{ \hat{D}_{\alpha}^{\mathrm{NN}} \}$ represents the subgroup consisting of the identity, 3 $C_2$ rotations, and the 8 $C_3$ rotations.

In this work, we have introduced the SSNN, a neural network architecture capable of discovering the entire discrete symmetry group in a physical system's parameter space. The SSNN comprises a standard neural network and a set of branches providing linear transformations. Upon training, the active branches yield the ground truth symmetry transformation matrices. We successfully applied this approach to various fields, demonstrating its wide applicability.  
It must be stressed that the SSNN method is self-sufficient, relying solely on a training dataset without any prior knowledge of the system's symmetry, the function $\vec{y}(\vec{x})$, or an oracle approximation of $\vec{y}$. This simplicity and flexibility may make it a powerful tool for symmetry analysis in diverse physical systems. Moreover, the SSNN can perform inverse design, finding the parameters that lead to specific magnitudes in the considered physical system.

We acknowledge Project PID2020-115221GB-C41, financed by MCIN/AEI/10.13039/501100011033, and the Aragon Government through Project Q-MAD. 
The code for this paper can be found online at \cite{github}.

\nocite{*}


\begin{thebibliography}{30}
\expandafter\ifx\csname natexlab\endcsname\relax\def\natexlab#1{#1}\fi
\expandafter\ifx\csname bibnamefont\endcsname\relax
  \def\bibnamefont#1{#1}\fi
\expandafter\ifx\csname bibfnamefont\endcsname\relax
  \def\bibfnamefont#1{#1}\fi
\expandafter\ifx\csname citenamefont\endcsname\relax
  \def\citenamefont#1{#1}\fi
\expandafter\ifx\csname url\endcsname\relax
  \def\url#1{\texttt{#1}}\fi
\expandafter\ifx\csname urlprefix\endcsname\relax\def\urlprefix{URL }\fi
\providecommand{\bibinfo}[2]{#2}
\providecommand{\eprint}[2][]{\url{#2}}

\bibitem[{\citenamefont{Noether}(1918)}]{Noether1918}
\bibinfo{author}{\bibfnamefont{E.}~\bibnamefont{Noether}},
  \bibinfo{journal}{Nachrichten von der Gesellschaft der Wissenschaften zu
  Göttingen, Mathematisch-Physikalische Klasse}
  \textbf{\bibinfo{volume}{1918}}, \bibinfo{pages}{235} (\bibinfo{year}{1918}),
  \urlprefix\url{http://eudml.org/doc/59024}.

\bibitem[{\citenamefont{Cohen-Tannoudji
  et~al.}(1977)\citenamefont{Cohen-Tannoudji, Diu, and
  Laloë}}]{Cohen-Tannoudji:101367}
\bibinfo{author}{\bibfnamefont{C.}~\bibnamefont{Cohen-Tannoudji}},
  \bibinfo{author}{\bibfnamefont{B.}~\bibnamefont{Diu}}, \bibnamefont{and}
  \bibinfo{author}{\bibfnamefont{F.}~\bibnamefont{Laloë}},
  \emph{\bibinfo{title}{{Quantum mechanics; 1st ed.}}}
  (\bibinfo{publisher}{Wiley}, \bibinfo{address}{New York, NY},
  \bibinfo{year}{1977}), \bibinfo{note}{trans. of : Mécanique quantique. Paris
  : Hermann, 1973}, \urlprefix\url{https://cds.cern.ch/record/101367}.

\bibitem[{\citenamefont{LeCun et~al.}(2015)\citenamefont{LeCun, Bengio, and
  Hinton}}]{LeCun_2015}
\bibinfo{author}{\bibfnamefont{Y.}~\bibnamefont{LeCun}},
  \bibinfo{author}{\bibfnamefont{Y.}~\bibnamefont{Bengio}}, \bibnamefont{and}
  \bibinfo{author}{\bibfnamefont{G.}~\bibnamefont{Hinton}},
  \bibinfo{journal}{Nature} \textbf{\bibinfo{volume}{521}},
  \bibinfo{pages}{436} (\bibinfo{year}{2015}),
  \urlprefix\url{https://doi.org/10.1038%2Fnature14539}.

\bibitem[{\citenamefont{Cybenko}(1989)}]{Cybenko1989}
\bibinfo{author}{\bibfnamefont{G.}~\bibnamefont{Cybenko}},
  \bibinfo{journal}{Mathematics of Control, Signals, and Systems}
  \textbf{\bibinfo{volume}{2}}, \bibinfo{pages}{303} (\bibinfo{year}{1989}),
  \urlprefix\url{https://doi.org/10.1007/bf02551274}.

\bibitem[{\citenamefont{Hornik}(1991)}]{Hornik1991}
\bibinfo{author}{\bibfnamefont{K.}~\bibnamefont{Hornik}},
  \bibinfo{journal}{Neural Networks} \textbf{\bibinfo{volume}{4}},
  \bibinfo{pages}{251} (\bibinfo{year}{1991}),
  \urlprefix\url{https://doi.org/10.1016/0893-6080(91)90009-t}.

\bibitem[{\citenamefont{Nielsen}(2015)}]{nielsen2015neural}
\bibinfo{author}{\bibfnamefont{M.}~\bibnamefont{Nielsen}},
  \emph{\bibinfo{title}{Neural Networks and Deep Learning}}
  (\bibinfo{publisher}{Determination Press}, \bibinfo{year}{2015}),
  \urlprefix\url{https://books.google.es/books?id=STDBswEACAAJ}.

\bibitem[{\citenamefont{Carleo et~al.}(2019)\citenamefont{Carleo, Cirac,
  Cranmer, Daudet, Schuld, Tishby, Vogt-Maranto, and
  Zdeborov\'a}}]{Carleo_2019}
\bibinfo{author}{\bibfnamefont{G.}~\bibnamefont{Carleo}},
  \bibinfo{author}{\bibfnamefont{I.}~\bibnamefont{Cirac}},
  \bibinfo{author}{\bibfnamefont{K.}~\bibnamefont{Cranmer}},
  \bibinfo{author}{\bibfnamefont{L.}~\bibnamefont{Daudet}},
  \bibinfo{author}{\bibfnamefont{M.}~\bibnamefont{Schuld}},
  \bibinfo{author}{\bibfnamefont{N.}~\bibnamefont{Tishby}},
  \bibinfo{author}{\bibfnamefont{L.}~\bibnamefont{Vogt-Maranto}},
  \bibnamefont{and}
  \bibinfo{author}{\bibfnamefont{L.}~\bibnamefont{Zdeborov\'a}},
  \bibinfo{journal}{Rev. Mod. Phys.} \textbf{\bibinfo{volume}{91}},
  \bibinfo{pages}{045002} (\bibinfo{year}{2019}),
  \urlprefix\url{https://link.aps.org/doi/10.1103/RevModPhys.91.045002}.

\bibitem[{\citenamefont{Liu et~al.}(2022)\citenamefont{Liu, Madhavan, and
  Tegmark}}]{Liu_2022}
\bibinfo{author}{\bibfnamefont{Z.}~\bibnamefont{Liu}},
  \bibinfo{author}{\bibfnamefont{V.}~\bibnamefont{Madhavan}}, \bibnamefont{and}
  \bibinfo{author}{\bibfnamefont{M.}~\bibnamefont{Tegmark}},
  \bibinfo{journal}{Phys. Rev. E} \textbf{\bibinfo{volume}{106}},
  \bibinfo{pages}{045307} (\bibinfo{year}{2022}),
  \urlprefix\url{https://link.aps.org/doi/10.1103/PhysRevE.106.045307}.

\bibitem[{\citenamefont{Liu and Tegmark}(2021)}]{Liu_2021}
\bibinfo{author}{\bibfnamefont{Z.}~\bibnamefont{Liu}} \bibnamefont{and}
  \bibinfo{author}{\bibfnamefont{M.}~\bibnamefont{Tegmark}},
  \bibinfo{journal}{Phys. Rev. Lett.} \textbf{\bibinfo{volume}{126}},
  \bibinfo{pages}{180604} (\bibinfo{year}{2021}),
  \urlprefix\url{https://link.aps.org/doi/10.1103/PhysRevLett.126.180604}.

\bibitem[{\citenamefont{Liu et~al.}(2023)\citenamefont{Liu, Sturm, Bharadwaj,
  Silva, and Tegmark}}]{https://doi.org/10.48550/arxiv.2305.19525}
\bibinfo{author}{\bibfnamefont{Z.}~\bibnamefont{Liu}},
  \bibinfo{author}{\bibfnamefont{P.~O.} \bibnamefont{Sturm}},
  \bibinfo{author}{\bibfnamefont{S.}~\bibnamefont{Bharadwaj}},
  \bibinfo{author}{\bibfnamefont{S.}~\bibnamefont{Silva}}, \bibnamefont{and}
  \bibinfo{author}{\bibfnamefont{M.}~\bibnamefont{Tegmark}},
  \emph{\bibinfo{title}{Discovering new interpretable conservation laws as
  sparse invariants}} (\bibinfo{year}{2023}), \eprint{2305.19525}.

\bibitem[{\citenamefont{Ha and Jeong}(2021)}]{Ha_2021}
\bibinfo{author}{\bibfnamefont{S.}~\bibnamefont{Ha}} \bibnamefont{and}
  \bibinfo{author}{\bibfnamefont{H.}~\bibnamefont{Jeong}},
  \bibinfo{journal}{Phys. Rev. Res.} \textbf{\bibinfo{volume}{3}},
  \bibinfo{pages}{L042035} (\bibinfo{year}{2021}),
  \urlprefix\url{https://link.aps.org/doi/10.1103/PhysRevResearch.3.L042035}.

\bibitem[{\citenamefont{Wetzel et~al.}(2020)\citenamefont{Wetzel, Melko, Scott,
  Panju, and Ganesh}}]{Wetzel_2020}
\bibinfo{author}{\bibfnamefont{S.~J.} \bibnamefont{Wetzel}},
  \bibinfo{author}{\bibfnamefont{R.~G.} \bibnamefont{Melko}},
  \bibinfo{author}{\bibfnamefont{J.}~\bibnamefont{Scott}},
  \bibinfo{author}{\bibfnamefont{M.}~\bibnamefont{Panju}}, \bibnamefont{and}
  \bibinfo{author}{\bibfnamefont{V.}~\bibnamefont{Ganesh}},
  \bibinfo{journal}{Phys. Rev. Res.} \textbf{\bibinfo{volume}{2}},
  \bibinfo{pages}{033499} (\bibinfo{year}{2020}),
  \urlprefix\url{https://link.aps.org/doi/10.1103/PhysRevResearch.2.033499}.

\bibitem[{\citenamefont{Greydanus et~al.}(2019)\citenamefont{Greydanus, Dzamba,
  and Yosinski}}]{https://doi.org/10.48550/arxiv.1906.01563}
\bibinfo{author}{\bibfnamefont{S.}~\bibnamefont{Greydanus}},
  \bibinfo{author}{\bibfnamefont{M.}~\bibnamefont{Dzamba}}, \bibnamefont{and}
  \bibinfo{author}{\bibfnamefont{J.}~\bibnamefont{Yosinski}}, in
  \emph{\bibinfo{booktitle}{Advances in Neural Information Processing
  Systems}}, edited by
  \bibinfo{editor}{\bibfnamefont{H.}~\bibnamefont{Wallach}},
  \bibinfo{editor}{\bibfnamefont{H.}~\bibnamefont{Larochelle}},
  \bibinfo{editor}{\bibfnamefont{A.}~\bibnamefont{Beygelzimer}},
  \bibinfo{editor}{\bibfnamefont{F.}~\bibnamefont{d\textquotesingle
  Alch\'{e}-Buc}}, \bibinfo{editor}{\bibfnamefont{E.}~\bibnamefont{Fox}},
  \bibnamefont{and} \bibinfo{editor}{\bibfnamefont{R.}~\bibnamefont{Garnett}}
  (\bibinfo{publisher}{Curran Associates, Inc.}, \bibinfo{year}{2019}),
  vol.~\bibinfo{volume}{32},
  \urlprefix\url{https://proceedings.neurips.cc/paper_files/paper/2019/file/26cd8ecadce0d4efd6cc8a8725cbd1f8-Paper.pdf}.

\bibitem[{\citenamefont{Cranmer et~al.}(2020)\citenamefont{Cranmer, Greydanus,
  Hoyer, Battaglia, Spergel, and
  Ho}}]{https://doi.org/10.48550/arxiv.2003.04630}
\bibinfo{author}{\bibfnamefont{M.}~\bibnamefont{Cranmer}},
  \bibinfo{author}{\bibfnamefont{S.}~\bibnamefont{Greydanus}},
  \bibinfo{author}{\bibfnamefont{S.}~\bibnamefont{Hoyer}},
  \bibinfo{author}{\bibfnamefont{P.}~\bibnamefont{Battaglia}},
  \bibinfo{author}{\bibfnamefont{D.}~\bibnamefont{Spergel}}, \bibnamefont{and}
  \bibinfo{author}{\bibfnamefont{S.}~\bibnamefont{Ho}},
  \emph{\bibinfo{title}{Lagrangian neural networks}} (\bibinfo{year}{2020}),
  \eprint{2003.04630}.

\bibitem[{\citenamefont{Lu et~al.}(2022)\citenamefont{Lu, Dangovski, and
  Soljačić}}]{https://doi.org/10.48550/arxiv.2208.14995}
\bibinfo{author}{\bibfnamefont{P.~Y.} \bibnamefont{Lu}},
  \bibinfo{author}{\bibfnamefont{R.}~\bibnamefont{Dangovski}},
  \bibnamefont{and}
  \bibinfo{author}{\bibfnamefont{M.}~\bibnamefont{Soljačić}},
  \emph{\bibinfo{title}{Discovering conservation laws using optimal transport
  and manifold learning}} (\bibinfo{year}{2022}), \eprint{2208.14995}.

\bibitem[{\citenamefont{Liu and Tegmark}(2022)}]{Liu_2022_Hid_Sym}
\bibinfo{author}{\bibfnamefont{Z.}~\bibnamefont{Liu}} \bibnamefont{and}
  \bibinfo{author}{\bibfnamefont{M.}~\bibnamefont{Tegmark}},
  \bibinfo{journal}{Phys. Rev. Lett.} \textbf{\bibinfo{volume}{128}},
  \bibinfo{pages}{180201} (\bibinfo{year}{2022}),
  \urlprefix\url{https://link.aps.org/doi/10.1103/PhysRevLett.128.180201}.

\bibitem[{\citenamefont{Bondesan and
  Lamacraft}(2019)}]{https://doi.org/10.48550/arxiv.1906.04645}
\bibinfo{author}{\bibfnamefont{R.}~\bibnamefont{Bondesan}} \bibnamefont{and}
  \bibinfo{author}{\bibfnamefont{A.}~\bibnamefont{Lamacraft}}, in
  \emph{\bibinfo{booktitle}{{ICML 2019 Workshop on Theoretical Physics for Deep
  Learning}}} (\bibinfo{year}{2019}), \eprint{1906.04645}.

\bibitem[{\citenamefont{Craven et~al.}(2022)\citenamefont{Craven, Croon,
  Cutting, and Houtz}}]{Craven_2022}
\bibinfo{author}{\bibfnamefont{S.}~\bibnamefont{Craven}},
  \bibinfo{author}{\bibfnamefont{D.}~\bibnamefont{Croon}},
  \bibinfo{author}{\bibfnamefont{D.}~\bibnamefont{Cutting}}, \bibnamefont{and}
  \bibinfo{author}{\bibfnamefont{R.}~\bibnamefont{Houtz}},
  \bibinfo{journal}{Phys. Rev. D} \textbf{\bibinfo{volume}{105}},
  \bibinfo{pages}{096030} (\bibinfo{year}{2022}),
  \urlprefix\url{https://link.aps.org/doi/10.1103/PhysRevD.105.096030}.

\bibitem[{\citenamefont{Forestano et~al.}(2023)\citenamefont{Forestano,
  Matchev, Matcheva, Roman, Unlu, and Verner}}]{Forestano_2023}
\bibinfo{author}{\bibfnamefont{R.~T.} \bibnamefont{Forestano}},
  \bibinfo{author}{\bibfnamefont{K.~T.} \bibnamefont{Matchev}},
  \bibinfo{author}{\bibfnamefont{K.}~\bibnamefont{Matcheva}},
  \bibinfo{author}{\bibfnamefont{A.}~\bibnamefont{Roman}},
  \bibinfo{author}{\bibfnamefont{E.~B.} \bibnamefont{Unlu}}, \bibnamefont{and}
  \bibinfo{author}{\bibfnamefont{S.}~\bibnamefont{Verner}},
  \bibinfo{journal}{Machine Learning: Science and Technology}
  \textbf{\bibinfo{volume}{4}}, \bibinfo{pages}{025027} (\bibinfo{year}{2023}),
  \urlprefix\url{https://dx.doi.org/10.1088/2632-2153/acd989}.

\bibitem[{\citenamefont{Krippendorf and Syvaeri}(2020)}]{Krippendorf_2020}
\bibinfo{author}{\bibfnamefont{S.}~\bibnamefont{Krippendorf}} \bibnamefont{and}
  \bibinfo{author}{\bibfnamefont{M.}~\bibnamefont{Syvaeri}},
  \bibinfo{journal}{Machine Learning: Science and Technology}
  \textbf{\bibinfo{volume}{2}}, \bibinfo{pages}{015010} (\bibinfo{year}{2020}),
  \urlprefix\url{https://dx.doi.org/10.1088/2632-2153/abbd2d}.

\bibitem[{\citenamefont{Yang et~al.}(2023)\citenamefont{Yang, Walters, Dehmamy,
  and Yu}}]{https://doi.org/10.48550/arxiv.2302.00236}
\bibinfo{author}{\bibfnamefont{J.}~\bibnamefont{Yang}},
  \bibinfo{author}{\bibfnamefont{R.}~\bibnamefont{Walters}},
  \bibinfo{author}{\bibfnamefont{N.}~\bibnamefont{Dehmamy}}, \bibnamefont{and}
  \bibinfo{author}{\bibfnamefont{R.}~\bibnamefont{Yu}},
  \emph{\bibinfo{title}{Generative adversarial symmetry discovery}}
  (\bibinfo{year}{2023}), \eprint{2302.00236}.

\bibitem[{\citenamefont{Dehmamy et~al.}(2021)\citenamefont{Dehmamy, Walters,
  Liu, Wang, and Yu}}]{NEURIPS2021_148148d6}
\bibinfo{author}{\bibfnamefont{N.}~\bibnamefont{Dehmamy}},
  \bibinfo{author}{\bibfnamefont{R.}~\bibnamefont{Walters}},
  \bibinfo{author}{\bibfnamefont{Y.}~\bibnamefont{Liu}},
  \bibinfo{author}{\bibfnamefont{D.}~\bibnamefont{Wang}}, \bibnamefont{and}
  \bibinfo{author}{\bibfnamefont{R.}~\bibnamefont{Yu}}, in
  \emph{\bibinfo{booktitle}{Advances in Neural Information Processing
  Systems}}, edited by
  \bibinfo{editor}{\bibfnamefont{M.}~\bibnamefont{Ranzato}},
  \bibinfo{editor}{\bibfnamefont{A.}~\bibnamefont{Beygelzimer}},
  \bibinfo{editor}{\bibfnamefont{Y.}~\bibnamefont{Dauphin}},
  \bibinfo{editor}{\bibfnamefont{P.}~\bibnamefont{Liang}}, \bibnamefont{and}
  \bibinfo{editor}{\bibfnamefont{J.~W.} \bibnamefont{Vaughan}}
  (\bibinfo{publisher}{Curran Associates, Inc.}, \bibinfo{year}{2021}),
  vol.~\bibinfo{volume}{34}, pp. \bibinfo{pages}{2503--2515},
  \urlprefix\url{https://proceedings.neurips.cc/paper_files/paper/2021/file/148148d62be67e0916a833931bd32b26-Paper.pdf}.

\bibitem[{\citenamefont{Desai et~al.}(2022)\citenamefont{Desai, Nachman, and
  Thaler}}]{Desai_2022}
\bibinfo{author}{\bibfnamefont{K.}~\bibnamefont{Desai}},
  \bibinfo{author}{\bibfnamefont{B.}~\bibnamefont{Nachman}}, \bibnamefont{and}
  \bibinfo{author}{\bibfnamefont{J.}~\bibnamefont{Thaler}},
  \bibinfo{journal}{Phys. Rev. D} \textbf{\bibinfo{volume}{105}},
  \bibinfo{pages}{096031} (\bibinfo{year}{2022}),
  \urlprefix\url{https://link.aps.org/doi/10.1103/PhysRevD.105.096031}.

\bibitem[{\citenamefont{Barenboim et~al.}(2021)\citenamefont{Barenboim, Hirn,
  and Sanz}}]{Barenboim_2021}
\bibinfo{author}{\bibfnamefont{G.}~\bibnamefont{Barenboim}},
  \bibinfo{author}{\bibfnamefont{J.}~\bibnamefont{Hirn}}, \bibnamefont{and}
  \bibinfo{author}{\bibfnamefont{V.}~\bibnamefont{Sanz}},
  \bibinfo{journal}{SciPost Phys.} \textbf{\bibinfo{volume}{11}},
  \bibinfo{pages}{014} (\bibinfo{year}{2021}),
  \urlprefix\url{https://scipost.org/10.21468/SciPostPhys.11.1.014}.

\bibitem[{\citenamefont{Decelle et~al.}(2019)\citenamefont{Decelle,
  Martin-Mayor, and Seoane}}]{Decelle_2019}
\bibinfo{author}{\bibfnamefont{A.}~\bibnamefont{Decelle}},
  \bibinfo{author}{\bibfnamefont{V.}~\bibnamefont{Martin-Mayor}},
  \bibnamefont{and} \bibinfo{author}{\bibfnamefont{B.}~\bibnamefont{Seoane}},
  \bibinfo{journal}{Phys. Rev. E} \textbf{\bibinfo{volume}{100}},
  \bibinfo{pages}{050102} (\bibinfo{year}{2019}),
  \urlprefix\url{https://link.aps.org/doi/10.1103/PhysRevE.100.050102}.

\bibitem[{\citenamefont{Zhang et~al.}(2018)\citenamefont{Zhang, Jin, Na, Zhang,
  and Yu}}]{Zhang_2018}
\bibinfo{author}{\bibfnamefont{C.}~\bibnamefont{Zhang}},
  \bibinfo{author}{\bibfnamefont{J.}~\bibnamefont{Jin}},
  \bibinfo{author}{\bibfnamefont{W.}~\bibnamefont{Na}},
  \bibinfo{author}{\bibfnamefont{Q.-J.} \bibnamefont{Zhang}}, \bibnamefont{and}
  \bibinfo{author}{\bibfnamefont{M.}~\bibnamefont{Yu}},
  \bibinfo{journal}{{IEEE} Transactions on Microwave Theory and Techniques}
  \textbf{\bibinfo{volume}{66}}, \bibinfo{pages}{3781} (\bibinfo{year}{2018}),
  \urlprefix\url{https://doi.org/10.1109%2Ftmtt.2018.2841889}.

\bibitem[{sup()}]{suppmat}
\bibinfo{note}{See Supplemental Material for the technical details of the
  calculations in (i) the n-th complex root problem, (ii) the nanophotonics
  problem and (iii) the quantum chemistry problem.}

\bibitem[{\citenamefont{Arfken et~al.}(2013)\citenamefont{Arfken, Weber, and
  Harris}}]{ARFKEN2013815}
\bibinfo{author}{\bibfnamefont{G.~B.} \bibnamefont{Arfken}},
  \bibinfo{author}{\bibfnamefont{H.~J.} \bibnamefont{Weber}}, \bibnamefont{and}
  \bibinfo{author}{\bibfnamefont{F.~E.} \bibnamefont{Harris}}, in
  \emph{\bibinfo{booktitle}{Mathematical Methods for Physicists (Seventh
  Edition)}}, edited by \bibinfo{editor}{\bibfnamefont{G.~B.}
  \bibnamefont{Arfken}}, \bibinfo{editor}{\bibfnamefont{H.~J.}
  \bibnamefont{Weber}}, \bibnamefont{and} \bibinfo{editor}{\bibfnamefont{F.~E.}
  \bibnamefont{Harris}} (\bibinfo{publisher}{Academic Press},
  \bibinfo{address}{Boston}, \bibinfo{year}{2013}), pp.
  \bibinfo{pages}{815--870}, \bibinfo{edition}{seventh edition} ed., ISBN
  \bibinfo{isbn}{978-0-12-384654-9},
  \urlprefix\url{https://www.sciencedirect.com/science/article/pii/B9780123846549000177}.

\bibitem[{\citenamefont{Newton}(1982)}]{Newton1982}
\bibinfo{author}{\bibfnamefont{R.~G.} \bibnamefont{Newton}},
  \emph{\bibinfo{title}{Scattering Theory of Waves and Particles}}
  (\bibinfo{publisher}{Springer Berlin Heidelberg}, \bibinfo{year}{1982}),
  \urlprefix\url{https://doi.org/10.1007/978-3-642-88128-2}.

\bibitem[{\citenamefont{Calvo-Barlés et~al.}(2023)\citenamefont{Calvo-Barlés,
  Rodrigo, Sánchez-Burillo, and Martín-Moreno}}]{github}
\bibinfo{author}{\bibfnamefont{P.}~\bibnamefont{Calvo-Barlés}},
  \bibinfo{author}{\bibfnamefont{S.~G.} \bibnamefont{Rodrigo}},
  \bibinfo{author}{\bibfnamefont{E.}~\bibnamefont{Sánchez-Burillo}},
  \bibnamefont{and}
  \bibinfo{author}{\bibfnamefont{L.}~\bibnamefont{Martín-Moreno}}
  (\bibinfo{year}{2023}),
  \urlprefix\url{https://github.com/pablocalvo7/Symmetry_Seeker_NN}.

\end{thebibliography}
\providecommand{\noopsort}[1]{}\providecommand{\singleletter}[1]{#1}%

\end{document}


\title{Supplemental Material}

\author{Pablo Calvo-Barl\'es}
\affiliation{Instituto de Nanociencia y Materiales de Arag\'on (INMA), CSIC-Universidad de Zaragoza, 50009 Zaragoza, Spain\looseness=-1}
\affiliation{Departamento de F\'isica de la Materia Condensada, Universidad de Zaragoza, 50009 Zaragoza, Spain\looseness=-1}

\author{Sergio G. Rodrigo}
\affiliation{Instituto de Nanociencia y Materiales de Arag\'on (INMA), CSIC-Universidad de Zaragoza, 50009 Zaragoza, Spain\looseness=-1}
\affiliation{Departamento de F\'isica Aplicada, Universidad de Zaragoza, 50009 Zaragoza, Spain\looseness=-1}

\author{Eduardo S\'anchez-Burillo}
\affiliation{PredictLand S.L., 50001 Zaragoza, Spain\looseness=-1}

\author{Luis Mart\'in-Moreno}
\email[]{lmm@unizar.es}
\affiliation{Instituto de Nanociencia y Materiales de Arag\'on (INMA), CSIC-Universidad de Zaragoza, 50009 Zaragoza, Spain\looseness=-1}
\affiliation{Departamento de F\'isica de la Materia Condensada, Universidad de Zaragoza, 50009 Zaragoza, Spain\looseness=-1}

\date{\today}

\maketitle



\section{n-th root problem}

In this section, we present the technical details of the SSNN training algorithm and data generation for the n-th complex root problem $w = z^n$. We extend our exploration to various root problems, including the complex third root (shown in the main text), the real second root, and the complex fourth and fifth roots. Additionally, we provide visual representations to enhance the understanding of the obtained activity values.

\begin{figure}
    \centering
    \includegraphics[scale=0.06]{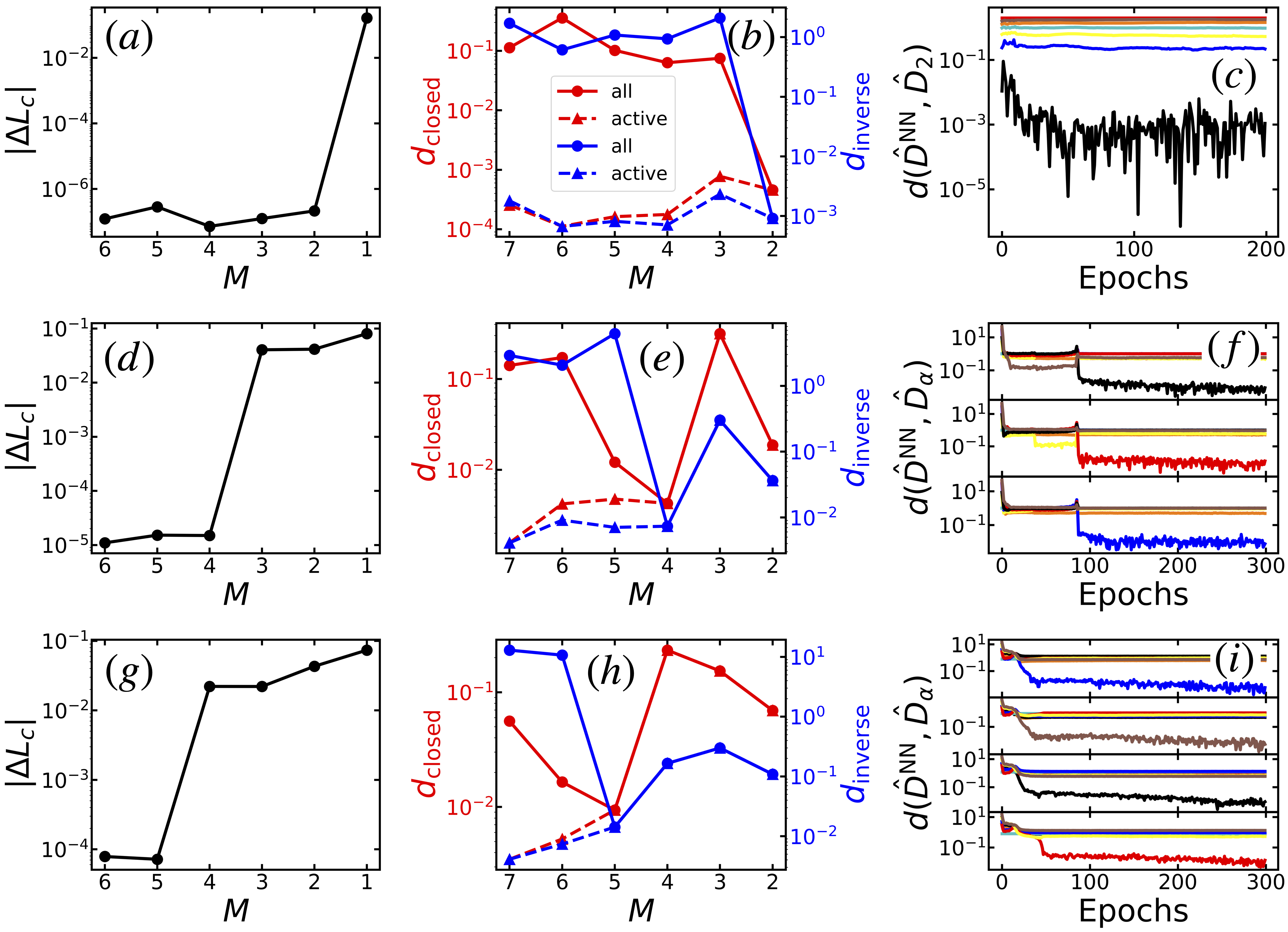}
    \caption{Symmetries in $w = z^{n}$: For each n-th root problem, we present three results: the loss variation as a function of the number of branches, the group metrics (for all branches and the active ones), and the distances between the predicted and ground truth symmetry matrices. Specifically, we show the real second root in (a-c), the complex fourth root in (d-f), and the complex fifth root in (g-i).}
    \label{fig:FIG_1}
\end{figure}

\subsubsection{SSNN training}

For all the NN calculations in this paper, we utilized Tensorflow \cite{tensorflow2015-whitepaper} and Keras \cite{chollet2015keras} libraries, with Adam \cite{kingma2017adam} as the gradient-descent optimizer for the loss function. Each n-th root problem was trained with 5000 training samples and 1000 validation samples. The $z$ values were randomly generated from a uniform distribution within the region $V = \{ z \in \mathbb{C} \; | \; |z| \leq 10 \}$ ($\mathbb{R}$ for the second root). Data was normalized as $z_i^{\mathrm{norm}} = z_i / 10$ and $w_i^{\mathrm{norm}} = z_i / 10^n$.

The standard NN architecture consisted of an input layer (2 neurons for the third, fourth, and fifth roots, and 1 neuron for the second), followed by two dense hidden layers with 10 neurons each using a sigmoid activation function. The latent layer had no activation function, only weights and biases. A learning rate of 0.01 and a mini-batch size of 128 samples were used. The SSNN was trained for 8000 epochs at each $M$-stage to ensure loss convergence. The maximum number of branches was $M_{\mathrm{max}}=7$ for each n-th root problem. Supplementary results for the second, fourth, and fifth roots can be observed in Fig. \ref{fig:FIG_1}.

\subsubsection{Branch activities}

\begin{figure}[b]
    \centering
    \includegraphics[scale=0.06]{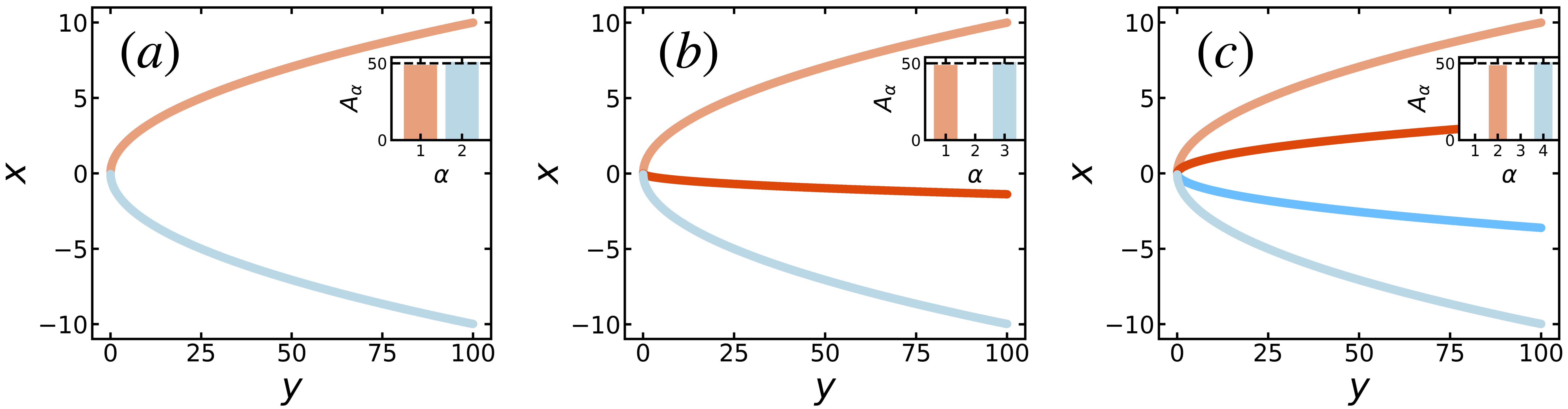}
    \caption{Visualization of predictions made by the active/inactive branches in the  second root problem. The plots display the branch predictions as a function of the training data $\{ y_i \}$, with each branch represented by a different color. The activities $A_{\alpha}$ for $\alpha = 1,..., M$ are shown in an inset using the same colors as the curves. (a) $M = 2$ branches. (b) $M=3$ branches. (c) $M=4$ branches. }
    \label{fig:FIG_2}
\end{figure}

We now show the visual representation of predictions and activities from active and inactive branches. Fig. \ref{fig:FIG_2} corresponds to the second root problem.  We represent, as function of the training input samples $\{ \vec{y}_i \}$, the predictions of each branch: $\{ \vec{x}^{(0)}_{\mathrm{NN}} (\vec{y}_i) \}, ..., \{ \vec{x}^{(\alpha)}_{\mathrm{NN}} (\vec{y}_i) \} ,..., \{ \vec{x}^{(M)}_{\mathrm{NN}} (\vec{y}_i) \}$. Three different $M$-stages are shown: $M=K$ (Fig. \ref{fig:FIG_2}a), $M=K+1$ (Fig. \ref{fig:FIG_2}b) and $M=K+2$ (Fig. \ref{fig:FIG_2}c). The inset displays the activities of the branches with the same color as the curves.  Active branches accurately fit the function $x = \pm \sqrt{y}$ throughout the full range of the parameter space $V$, used in the training,  while inactive branches do not provide a correct approximation, consistent with the $A_{\alpha}$ values.  This is in accordance with what is expected from the $A_{\alpha}$ values.

\begin{figure}
    \centering
    \includegraphics[scale=0.06]{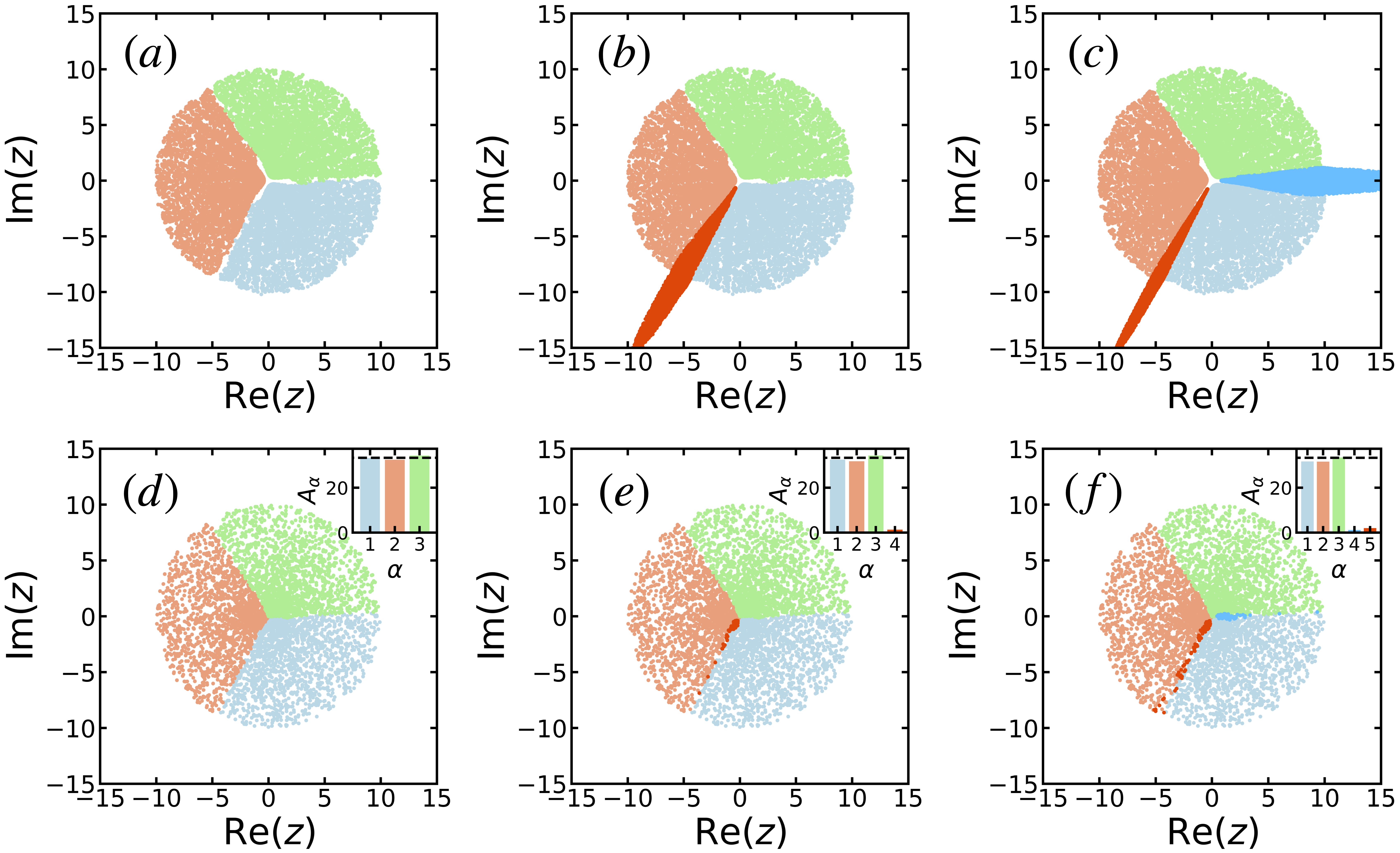}
    \caption{Visualization of predictions made by the active/inactive branches in the  third root problem. The upper row plots display the branch predictions in the parameter space, with each branch represented by a different color. The lower row plots show the output training samples$\{ \vec{x}_i \}$ in the parameter space, where each color corresponds to the branch with the minimum-MSE prediction. An inset shows the activities $A_{\alpha}$ for $\alpha = 1,..., M$ with the same colors as the points. (a,d) $M = 3$ branches. (b,e) $M=4$ branches. (c,f) $M=5$ branches.}
    \label{fig:FIG_3}
\end{figure}

\begin{figure}
    \centering
    \includegraphics[scale=0.06]{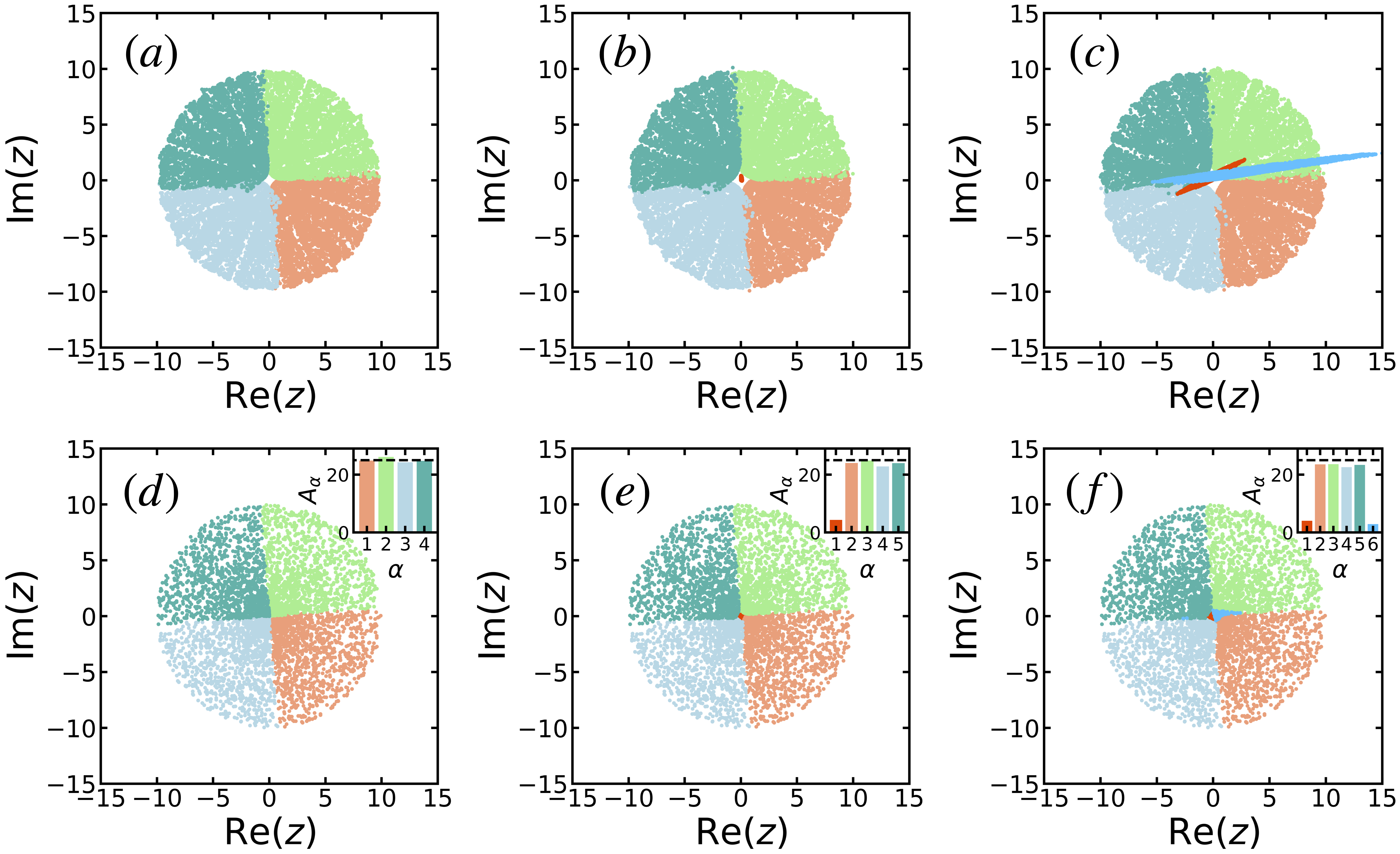}
    \caption{Visualization of predictions made by the active/inactive branches in the  fourth root problem. (a,d) $M = 4$ branches. (b,e) $M=5$ branches. (c,f) $M=6$ branches.}
    \label{fig:FIG_4}
\end{figure}

\begin{figure}
    \centering
    \includegraphics[scale=0.06]{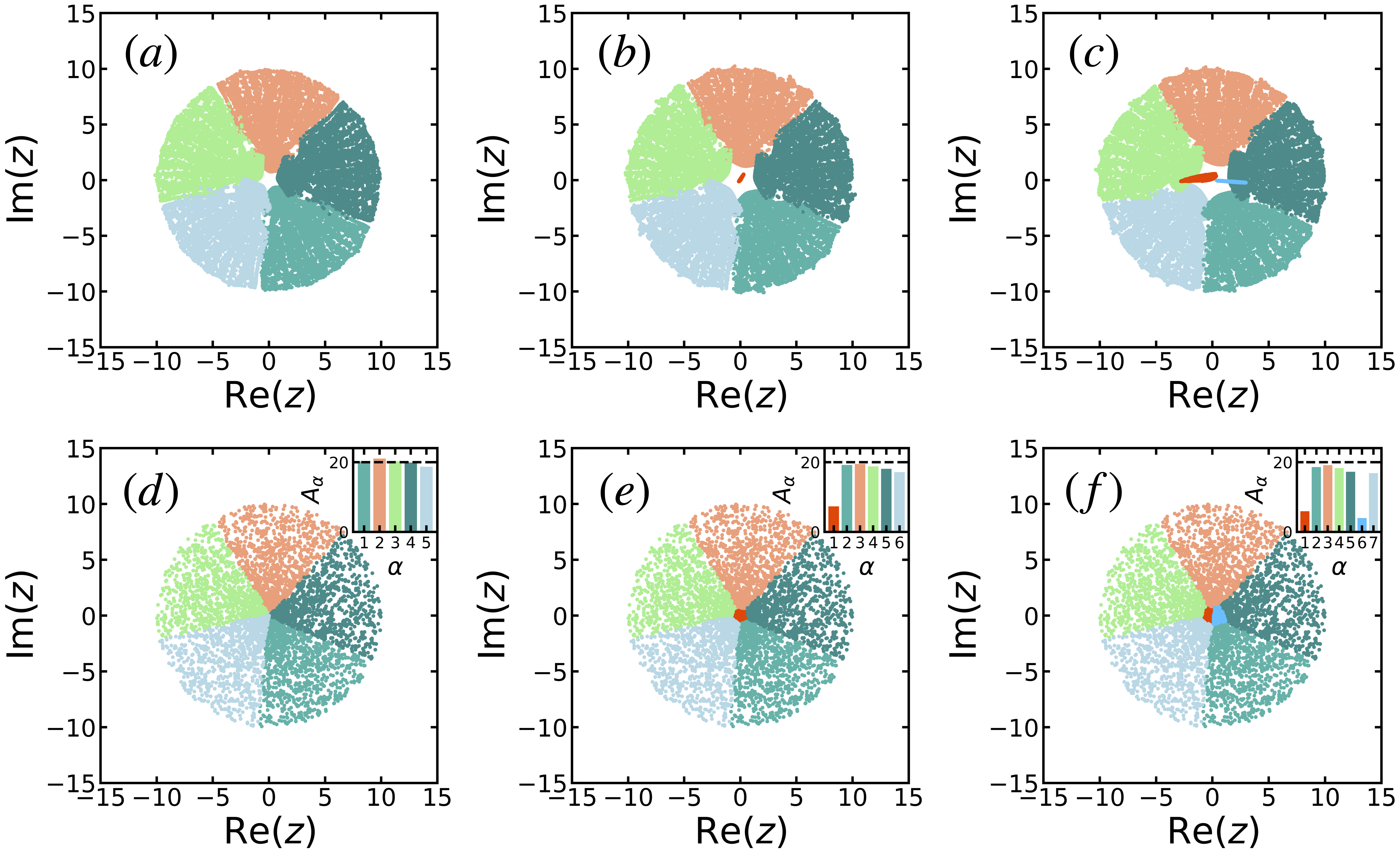}
    \caption{Visualization of predictions made by the active/inactive branches in the  fifth root problem. (a,d) $M = 5$ branches. (b,e) $M=6$ branches. (c,f) $M=7$ branches.}
    \label{fig:FIG_5}
\end{figure}

Figs. \ref{fig:FIG_3}, \ref{fig:FIG_4} and \ref{fig:FIG_5} correspond to the third, fourth and fifth root, respectively. In these figures, we only represent the parameter space (complex plane $z$).  Two types of scatter plots are shown for each training stage $M=K$, $M=K+1$ and $M=K+2$. The first type (upper row in the figures) displays $\{ \vec{x}^{(\alpha)}_{\mathrm{NN}} (\vec{y}_i) \}$ for $\alpha = 1,...,M$.  As expected, the active branches cover the irreducible regions maximally. However, small regions in $V$, particularly near the boundaries between irreducible regions, are not fully covered by the active branches. Notice that, close to boundaries, two branches provide predictions with similar losses.  This makes the inactive branches specialize in covering these small regions, explaining why $A_{\alpha}$ is not exactly zero for the inactive branches. The second type (lower row) plots the output samples $\{ \vec{x}_i \}$ from the training data set (uniformly distributed). The color of each point represents the branch with the minimum MSE prediction for that sample, indicating the area of the parameter space covered by each branch, directly related to its activity. The corresponding activities are shown in the insets.

\newpage

\section{Optical transmission in multilayer stack}

In this section, we outline the calculation details of the optical transmission problem discussed in the main text. First, we show how to calculate the transmission spectrum from a given set of dielectric materials. Subsequently, we present the SSNN training specifics, along with the convergence of the predicted symmetry matrices to the ground truth matrices.
\subsubsection{Transmission spectrum calculation}

\begin{figure}
    \centering
    \includegraphics[scale=0.04]{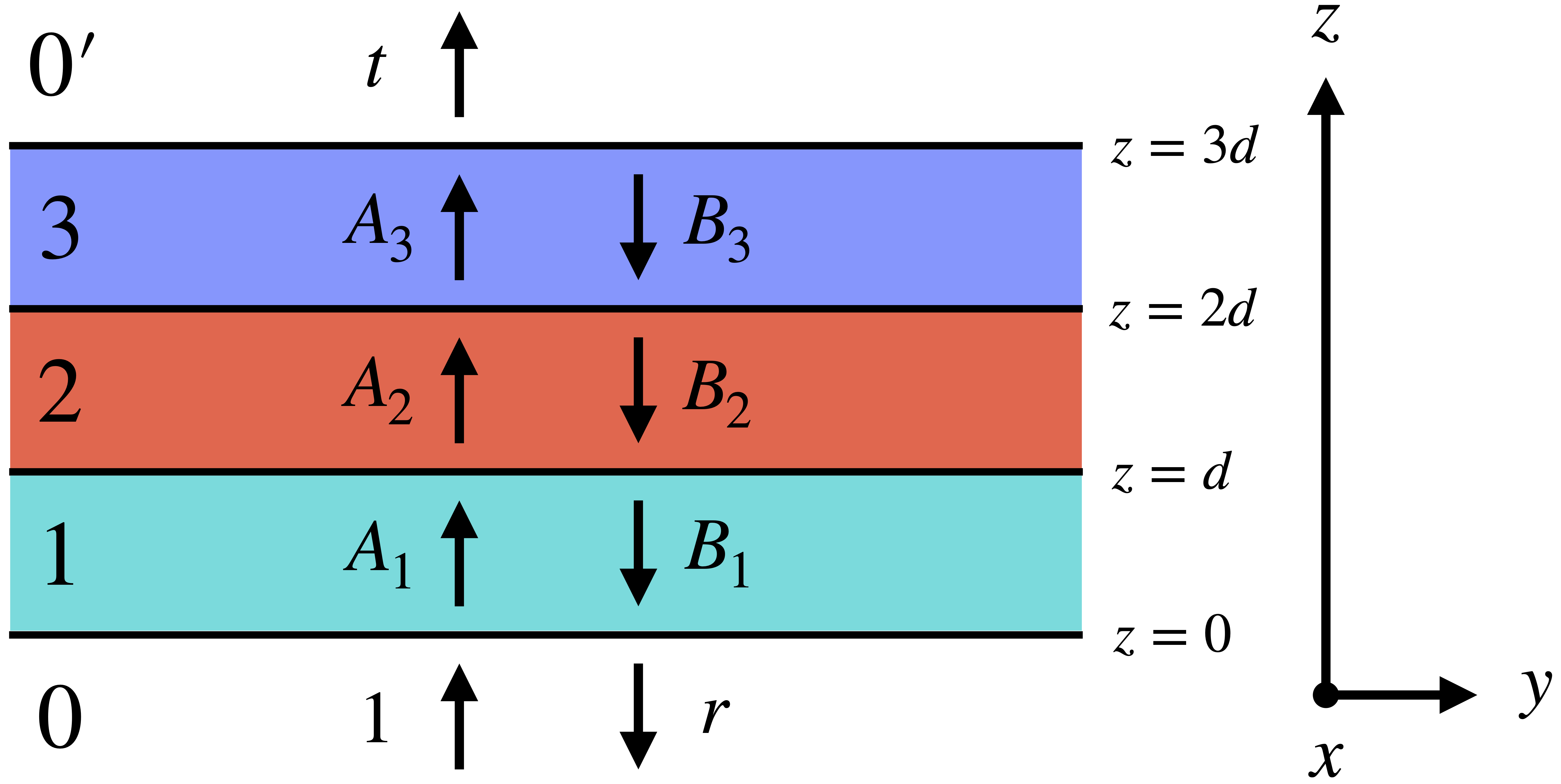}
    \caption{Schematic representation of the dielectric multilayer stack. The direction of light propagation is along the $z$-axis. The amplitudes of the transmitted modes are denoted as $A_j$, while the amplitudes of the reflected modes are denoted as $B_j$. The transmission coefficient is represented by $t$, and the reflection coefficient by $r$.}
    \label{fig:FIG_6}
\end{figure}

We study a photonic system described in the third section of the main text.  The system consists of three stacked layers (labeled as 1, 2, and 3) with the same width $d$ and dielectric constants $\varepsilon_1$, $\varepsilon_2$, and $\varepsilon_3$ (see Fig. \ref{fig:FIG_6}). In our calculations, we set $d = 250$ nm, and $\varepsilon_i$ varies within the range $[2,5]$. Media $0$ and $0'$ are treated as vacuum ($\varepsilon_0 = \varepsilon_{0'} = 1$).

An incident electromagnetic wave incident, with wavelength $\lambda$, impinging normally into the system.  This incident wave is transmitted and reflected at each interface. We express the magnitude of the electric field in each medium as follows:
\begin{equation}
   E_0 = e^{ikz} + re^{-ikz},
\end{equation}
\begin{equation}
   E_j = A_j e^{i k_j \left( z-(j-1)d \right) } + B_j e^{-i k_j \left( z-(j-1)d \right) },
\end{equation}
\begin{equation}
   E_{0'} = t e^{ik(z-3d)}.
\end{equation}
In the expressions above, $A_j$ and $B_j$ are the complex amplitudes of upgoing and downgoing waves, respectively,  at each medium $j$, while $t$ and $r$ are are the complex transmission and reflection coefficients of the whole system. Besides, $k = 2 \pi / \lambda$ is the wavevector at the vacuum and $k_j = \sqrt{\varepsilon_j}k$ is the wavevector at the medium $j$.  Additionally, $k = 2\pi/\lambda$ denotes the wavevector in vacuum, and $k_j = \sqrt{\varepsilon_j}k$ is the wavevector in medium $j$. Using these electric field magnitudes, we can calculate the magnetic field modules $H_j$ by applying the following relations derived from Maxwell's equations:
\begin{equation}
   -\vec{u}_z \times \vec{H_j} = \pm y_j \vec{E_j},
\end{equation}
In the expressions above, $\vec{u}_z$ represents the unitary vector in the $z$ direction, and $y_j = \sqrt{\varepsilon_j}$ is the modal admittance. The positive sign is used for the transmitted modes, and the negative sign is used for the reflected modes. Once we have obtained the electric and magnetic fields, we apply the boundary conditions at each interface:
\begin{equation}
   E_{j}(jd) = E_{j+1}(jd),
\end{equation}
\begin{equation}
   H_{j}(jd) = H_{j+1}(jd),
\end{equation}
for $j = 0,1,2,3$. We then obtain a linear system of 8 equations with 8 unknowns, $r, A_1, B_1, A_2, B_2, A_3, B_3$ and $t$:
\begin{equation}
\begin{cases}
   r - A_1 - B_1 = -1 \\
   -r - y_1 A_1 + y_1 B_1 = -1 \\
   A_1 e^{i k_1 d} + B_1 e^{-i k_1 d} - A_2 - B_2 = 0 \\
   y_1 e^{i k_1 d} A_1 - y_1 e^{-i k_1 d} B_1 - y_2 A_2 + y_2 B_2 = 0 \\
   e^{i k_2 d} A_2 + e^{-i k_2 d} B_2 - A_3 - B_3 = 0 \\
   y_2 e^{i k_2 d} A_2 - y_2 e^{-i k_2 d} B_2 - y_3 A_3 + y_3 B_3 = 0 \\
   -t + e^{i k_3 d} A_3 + e^{-i k_3 d} B_3 = 0 \\
   -t + y_3 e^{i k_3 d} A_3 - y_3 e^{-i k_3 d} B_3 = 0.
\end{cases}
\end{equation}

\begin{figure}
    \centering
    \includegraphics[scale=0.4]{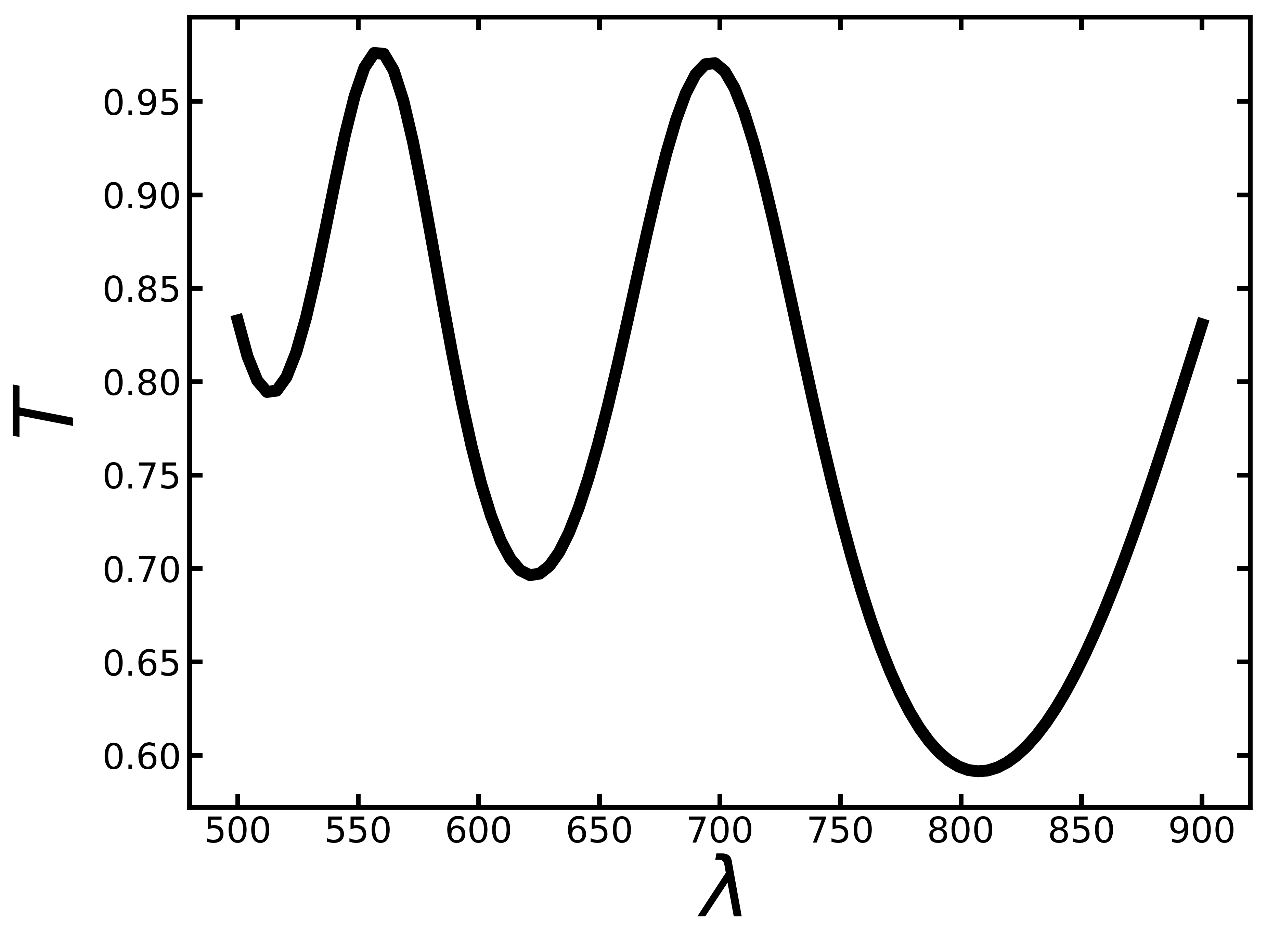}
    \caption{Example of the transmittance spectrum. The corresponding dielectric constants of the materials are $\varepsilon_1 = 2.5$, $\varepsilon_2 = 3.5$ and $\varepsilon_3 = 4.8$.}
    \label{fig:FIG_7}
\end{figure}

To calculate the transmission coefficient $t$ for a given wavelength $\lambda$, we solve this linear system.  After this, the transmittance is obtained as denoted as $T(\lambda) = |t|^2$. For the SSNN training data set, we calculate transmittance spectra by considering a discretized set of $m = 100$ equidistant wavelengths between $\lambda_{\mathrm{min}} = 500$ nm and $\lambda_{\mathrm{max}} = 900$ nm. An illustrative example is presented in Fig. \ref{fig:FIG_7}. 

\subsubsection{SSNN training}

We have trained the SSNN with 9000 training samples and 1000 validation samples. For the SSNN, we have normalized the input samples $\vec{\varepsilon}_i$ and output samples $\vec{T}_i$ as follows: $\vec{\varepsilon}_i^{\mathrm{norm}} = \left( \vec{\varepsilon}_i - (2,2,2)^T \right)/3$ and $\vec{T}_i^{\mathrm{norm}} = \left( \vec{T}_i - (T_{\mathrm{min}}, ..., T_{\mathrm{min}})^T \right) /(T_{\mathrm{max}} - T_{\mathrm{min}})$, being $T_{\mathrm{max}}$ and $T_{\mathrm{min}}$ the highest and lowest transmission values in the training data set, respectively.  
The standard section of the SSNN architecture consists of an input layer with $m = 100$ neurons and two non-linear hidden layers, each with 200 neurons and sigmoid activation function. The latent layer is linear (i.e., we have not used an activation function). During training, we used a learning rate of 0.01 and a mini-batch size of 100 samples. The SSNN was trained for 2000 epochs for each $M$-stage, and the maximum number of branches was set to $M_{\mathrm{max}}=7$.

\begin{figure}
    \centering
    \includegraphics[scale=0.45]{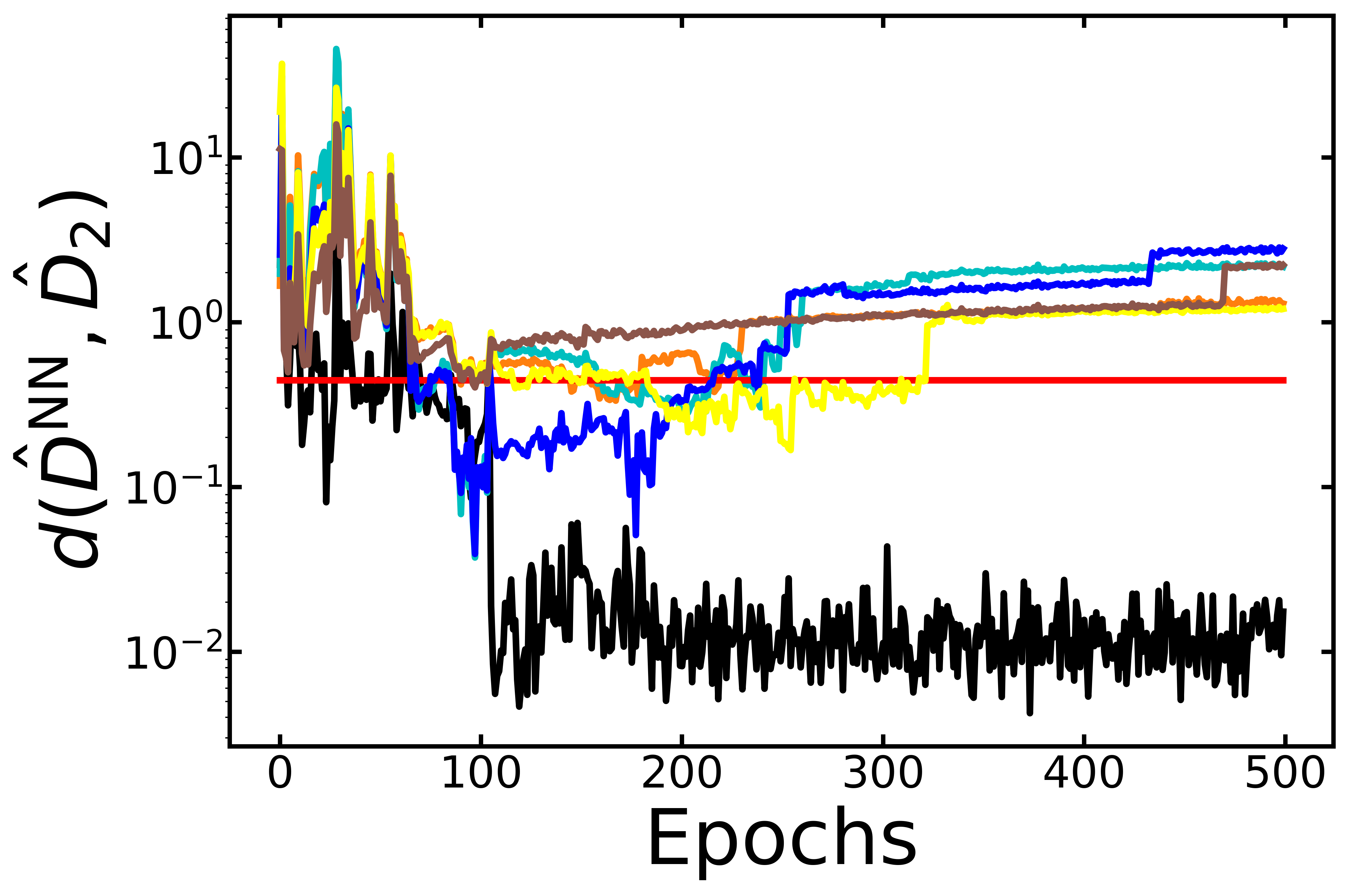}
    \caption{Evolution of the distance between the predicted matrices and the ground truth matrix $\hat{D}_2$ during the first training stage (M = 7) in the optical transmission problem.}
    \label{fig:FIG_8}
\end{figure}

In Fig. \ref{fig:FIG_8}, we present the evolution of matrix distances between each learned matrix $\hat{D}_{\alpha}^{\mathrm{NN}}$ and the inversion transformation in the order of the dielectric constants $\hat{D}_2$ (a symmetry in parameter space originating from the time reversal ) during the first training stage ($M = 7$). Interestingly, only one curve decays to $10^{-2}$. 

\newpage

\section{ENERGY SPECTRUM IN MOLECULES}

In this final section, we provide the calculation details of the molecule eigenspectrum problem as discussed in the main text. Firstly, we outline the method to calculate the eigenenergies of each molecular system. Subsequently, we present the SSNN training details, including the network hyperparameters and the branch activities at different $M$-stages. Finally, we demonstrate the convergence of the predicted symmetry matrices to the ground truth ones for each molecule.

\subsubsection{Eigenvalues calculation}
We consider the three types of 4-atom molecules introduced in the main text: rectangular, squared, and tetrahedral.  We describe these systems with a tight-binding model with one atomic level per atom. Hopping is assumed to occur only between nearest neighbors. The Hamiltonians of each system are:
\begin{equation}
\hat{H}_R = \begin{pmatrix}
e_1 & J_x & 0 & J_y\\
J_x & e_2 & J_y & 0\\
0 & J_y & e_3 & J_x\\
J_y & 0 & J_x & e_4
\end{pmatrix},
\end{equation}
\begin{equation}
\hat{H}_S = \begin{pmatrix}
e_1 & J & 0 & J\\
J & e_2 & J & 0\\
0 & J & e_3 & J\\
J & 0 & J & e_4
\end{pmatrix},
\end{equation}
\begin{equation}
\hat{H}_T = \begin{pmatrix}
e_1 & J & J & J\\
J & e_2 & J & J\\
J & J & e_3 & J\\
J & J & J & e_4
\end{pmatrix},
\end{equation}
where $\{ e_i \}$ are the atomic-energies and $J , J_x , J_y$ are the hopping terms. The eigenvalues ${ E_i }$ are obtained by diagonalizing these matrices. 

For the training data set, we generate random uniform atomic energies within specified ranges, while fixing the values of the hopping terms. In the rectangular molecule, we take $e_i \in [1,30]$, $J_x = 10.0$, and $J_y = 7.62$. In the square molecule, $e_i$ is within the range $[10,20]$, and $J$ is set to $10.0$. In the tetrahedral molecule, $e_i$ is within $[10,20]$ and $J$ is set to $1.0$.

\subsubsection{SSNN training}

\begin{figure}
    \centering
    \includegraphics[scale=0.45]{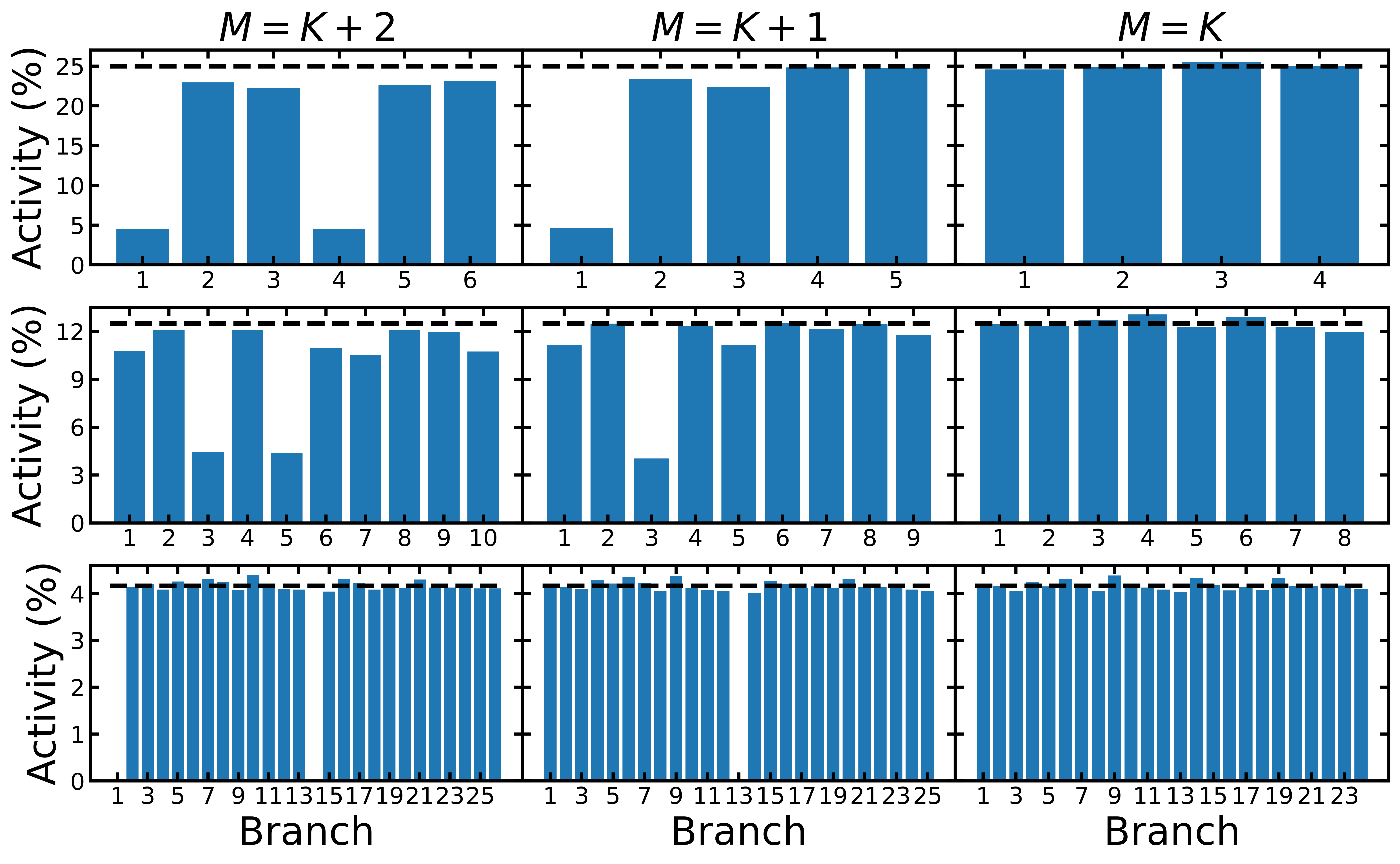}
    \caption{Branch activities $A_{\alpha}$ for $\alpha = 1,...,M$ at three different stages: $M = K+2$, $M = K+1$ and $M = K$. Each row corresponds to a different molecule: the upper row corresponds to the rectangular molecule, the middle row to the square molecule, and the lower row to the tetrahedral molecule.}
    \label{fig:FIG_9}
\end{figure}

For each molecule, we have trained the SSNN with 70000 training samples and 30000 validation samples. The standard section of the SSNN is composed by an input layer with $m = 4$ neurons and 2 non-linear hidden layers with 100 neurons each, using sigmoid activation functions. The latent layer is linear.  The learning rate was set to 0.01, and a mini-batch size of 200 samples was used.  We trained the SSNN for 3000 epochs for each $M$-stage. The maximum number of branches was set to $M_{\mathrm{max}}=8$ for the rectangle, $M_{\mathrm{max}}=12$ for the square, and $M_{\mathrm{max}}=30$ for the tetrahedron.

Fig. \ref{fig:FIG_9} displays the branch activities for each molecular system at three different stages: $M = K+2$, $M = K+1$ and $M = K$. In all cases, the values $A_{\alpha}$ indicate the presence of $K$ active branches and $M-K$ inactive branches.

\begin{figure}
    \centering
    \includegraphics[scale=0.4]{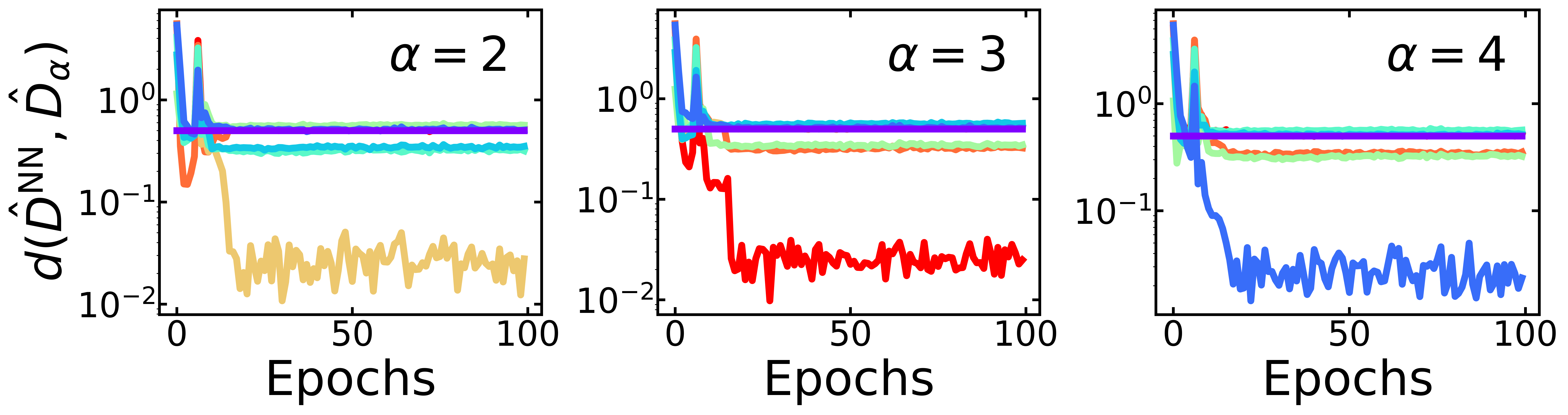}
    \caption{Evolution of the distance between the predicted matrices and the ground truth ones during the first training stage (M = 8) for the rectangular molecule. Each panel  corresponds to a specific ground truth symmetry  $\hat{D}_{\alpha}$.}
    \label{fig:FIG_10}
\end{figure}

\begin{figure}
    \centering
    \includegraphics[scale=0.3]{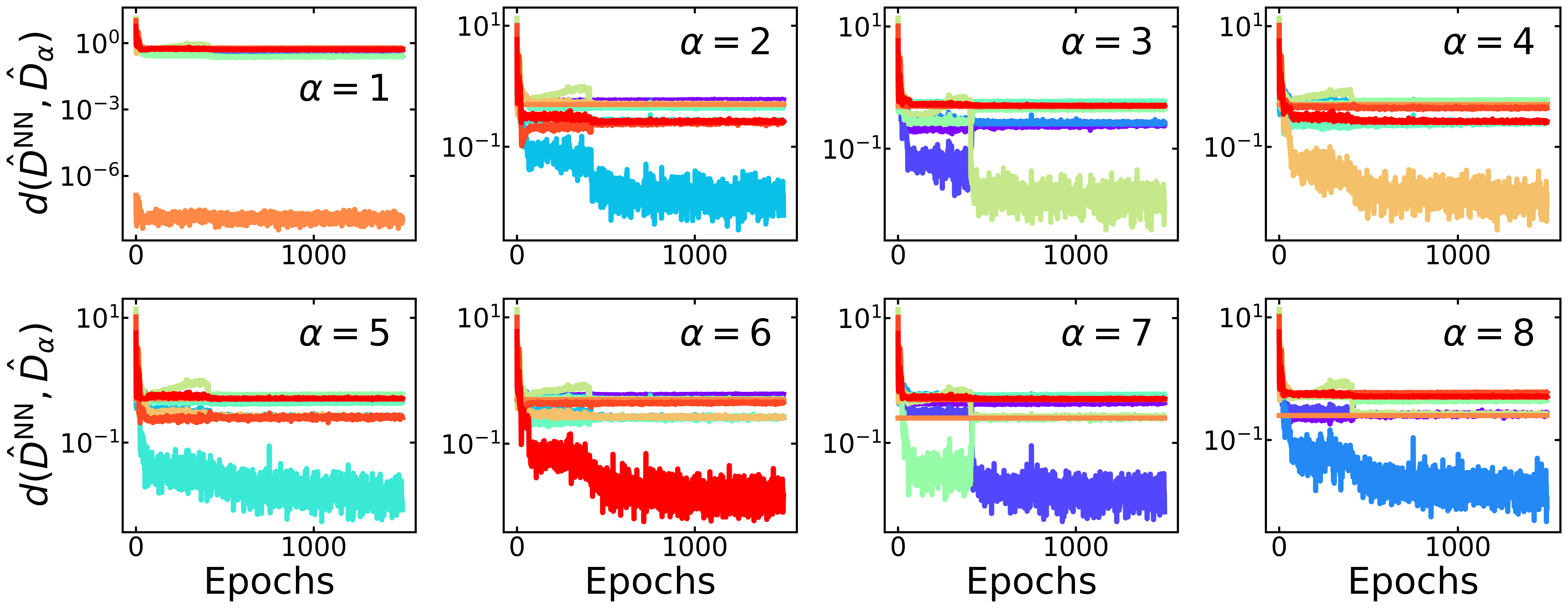}
    \caption{Evolution of the distance between the predicted matrices and the ground truth ones during the first training stage (M = 12) for the square molecule. Each panel  corresponds to a specific ground truth symmetry $\hat{D}_{\alpha}$.}
    \label{fig:FIG_11}
\end{figure}

\begin{figure}
    \centering
    \includegraphics[scale=0.7]{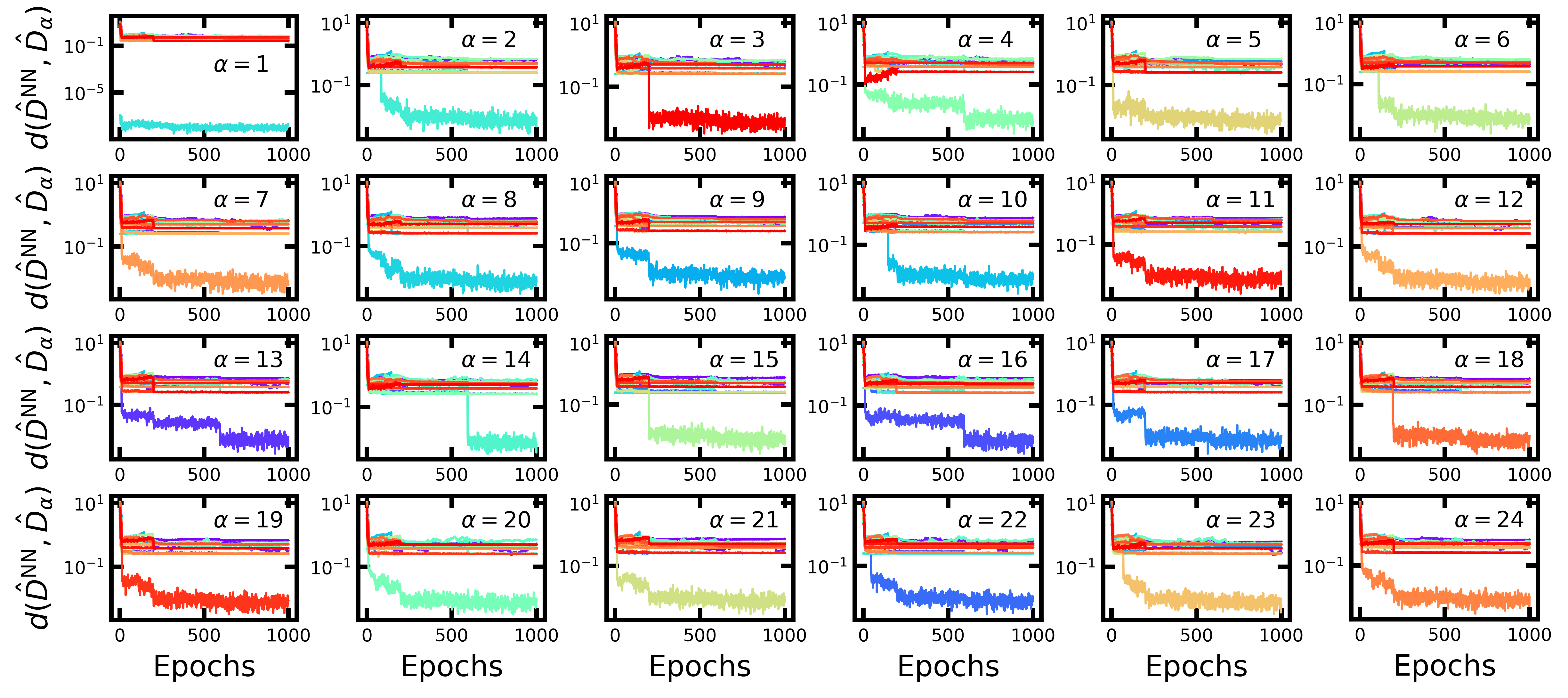}
    \caption{Evolution of the distance between the predicted matrices and the ground truth ones during the first training stage (M = 30) for the tetrahedral molecule. Each panel corresponds to a specific ground truth symmetry $\hat{D}_{\alpha}$.}
    \label{fig:FIG_12}
\end{figure}

Figs. \ref{fig:FIG_10}, \ref{fig:FIG_11} and \ref{fig:FIG_12} depict the evolution of the distance  between each learned matrix $\hat{D}_{\alpha}^{\mathrm{NN}}$ and each ground truth transformation $\hat{D}_{\beta}$ during the first training stage $M = M_{\mathrm{max}}$. As expected, these plots demonstrate that each ground truth symmetry matrix is discovered by a different active branch.

\nocite{*}

\newpage

\providecommand{\noopsort}[1]{}\providecommand{\singleletter}[1]{#1}%